\documentclass[apj]{emulateapj}

\newcommand{\ud}{\mathrm{d}}

\shorttitle{Giant H~II Region in M101}
\shortauthors{Sun et al.}

\begin{document}


\title{Giant H~II Regions in M101. I. X-ray Analysis of Hot Gas}






\author{Wei~Sun\altaffilmark{1},
        Yang~Chen\altaffilmark{1,2,6},
        Li~Feng\altaffilmark{1},
        You-Hua~Chu\altaffilmark{3},
        C.-H.~Rosie~Chen\altaffilmark{4},
        Q.~Daniel~Wang\altaffilmark{5},
	Jiang-Tao~Li\altaffilmark{2,5}}
\altaffiltext{1}{Department of Astronomy, Nanjing University, 
                 Nanjing 210093, China}
\altaffiltext{2}{Key Laboratory of Modern Astronomy and Astrophysics, 
                 Nanjing University, Ministry of Education, China}
\altaffiltext{3}{Department of Astronomy, University of Illinois, 
                 1002 West Green Street, Urbana, IL 61801}
\altaffiltext{4}{Max Planck Institut f{\"u}r Radioastronomie, 
                 Auf dem H{\"u}gel 69, 53121 Bonn, Germany}
\altaffiltext{5}{Department of Astronomy, University of Massachusetts, 
                 Amherst, MA 01003, USA}
\altaffiltext{6}{Author to whom any correspondence should be addressed.}


\begin{abstract}
We performed a \emph{Chandra} X-ray study of three giant H~II regions 
(GHRs), NGC~5461, NGC~5462, and NGC~5471, in the spiral galaxy M101.
The X-ray spectra of the three GHRs all contain a prominent thermal 
component with a temperature of $\sim0.2$~keV. In NGC~5461, the spatial 
distribution of the soft ($<1.5$~keV) X-ray emission is generally in 
agreement with the extent of H1105, the most luminous H~II region therein, 
but extends beyond its southern boundary, which could be attributed 
to outflows from the star cloud between H1105 and H1098. In NGC~5462, 
the X-ray emission is displaced from the \ion{H}{2} regions and a ridge 
of blue stars; the H$\alpha$ filaments extending from the ridge of 
star cloud to the diffuse X-rays suggest that hot gas outflows have 
occured. The X-rays from NGC~5471 are concentrated at the 
B-knot, a ``hypernova remnant'' candidate. Assuming a Sedov-Taylor 
evolution, the derived explosion energy, 
on the order of $10^{52}$~ergs, is consistent with a hypernova origin. 
In addition, a bright source in the field of NGC~5462 has been identified 
as a background AGN, instead of a black hole X-ray binary in M101.
\end{abstract}


\section{Introduction}
Intense star formation and subsequent evolution of massive stars are 
spectacular processes observable from radio through optical to X-ray and 
even $\gamma$-ray wavelengths, and represent a microcosm of starburst 
astrophysics. Upon formation from giant molecular clouds, young massive 
stars emit large amounts of ultraviolet (UV) radiation to ionize the ambient 
medium into H~II regions. The accumulated kinetic energy injected by stellar
winds from O-type and Wolf-Rayet stars rival the explosion energies of 
supernovae (SNe). These mechanical energies not only shock-heat the ambient 
medium to 10$^6$--10$^8$~K that emits diffuse X-rays, but also may compress 
ambient clouds to trigger further star formation. Within this medium, newly 
formed X-ray binaries and magnetized young stellar objects shine as point 
sources in X-rays. Regions of massive star formation are thus excellent 
astrophysical laboratories to study the co-evolution of massive stars and 
the multi-phase interstellar medium. 

Giant H~II regions (GHRs) are sites of intense star formation qualified as 
starbursts. Their H$\alpha$ luminosities, typically 
$\sim{}10^{39}$-$10^{41}$~erg~s$^{-1}$ \citep{1984ApJ...287..116K}, require 
an ionizing power equivalent to that of 24-2400 O5V stars 
\citep{1997A&A...322..598S}. 
Depending on their stellar content and 
interstellar environments, GHRs may possess very different X-ray properties. 
30 Doradus (30~Dor) in the Large Magellanic Cloud (LMC) is dominated by a 
central massive cluster, R136 \citep{1960MNRAS.121..337F} with a mass of 
$\sim{}6\times10^4\,M_\odot$ \citep{2005ASSL..329...49B}, and its diffuse 
X-ray luminosity is $L_{\rm X}\simeq1.4\times10^{37}$~erg~s$^{-1}$ in 
0.5--8~keV \citep{1999ApJ...510L.139W, 2006AJ....131.2140T}.
NGC~604 in M33 contains multiple OB associations spreading over a 
large area \citep{1996ApJ...456..174H}, and its hot ionized medium (HIM) has 
a total unabsorbed X-ray luminosity of $1.43\times10^{36}$~erg~s$^{-1}$ 
\citep{2008ApJ...685..919T}.
IC~131, another GHR in M33, shows diffuse X-ray emission within a large 
shell southeast to a concentration of OB stars and the diffuse X-ray emission 
is characterized by an unusually hard spectrum \citep{2009ApJ...707.1361T}.

\begin{deluxetable*}{cccccccc}[!htp]
\tabletypesize{\scriptsize}
\tablecaption{Journal of analyzed \emph{Chandra} ACIS Observation\label{tab3}}
\tablewidth{0pt}
\tablehead{
\colhead{ObsID} & \colhead{Start Date} & \colhead{Exposure (ks)\tablenotemark{a}} & 
\multicolumn{4}{c}{CCD chips\tablenotemark{b}} & 
\colhead{}    \\
\cline{4-7}\\
\colhead{} & \colhead{} & \colhead{} & \colhead{NGC~5461}& 
\colhead{NGC~5462} & \colhead{NGC~5471} & \colhead{Src.\ 6\tablenotemark{c}} &
\colhead{Data Mode}
}
\startdata
2779 &02.Oct.31 &11.9 &\nodata &\nodata &S3      &S2      & VFAINT\\
\hline
\noalign{\smallskip}
2065 &00.Oct.29 &9.63 &\nodata &S2      &S2      &S2      & FAINT\\
4737 &04.Jan.01 &20.5 &S3      &S3      &\nodata &S3      & VFAINT\\
4731 &04.Jan.19 &55.8 &\nodata &\nodata &S2      &\nodata & VFAINT\\
5297 &04.Jan.25 &18.7 &\nodata &\nodata &S2      &\nodata & VFAINT\\
5322 &04.May.03 &64.7 &\nodata &\nodata &\nodata &S2      & VFAINT\\
4733 &04.May.07 &23.6 &\nodata &\nodata &\nodata &S2      & VFAINT\\
5323 &04.May.09 &42.5 &\nodata &\nodata &\nodata &S2      & VFAINT\\
4736 &04.Nov.01 &73.9 &S2      &S2      &S2      &S2      & VFAINT\\
6152 &04.Nov.07 &22.7 &S2      &S2      &S2      &S2      & VFAINT\\
6170 &04.Dec.22 &43.4 &S3      &S3      &S2      &S3      & VFAINT\\
6175 &04.Dec.24 &40.6 &S3      &S3      &S2      &S3      & VFAINT\\
6169 &04.Dec.30 &28.0 &S3      &S3      &\nodata &S3      & VFAINT\\
\enddata
\tablenotetext{a}{Effective exposure, in which deadtime and bad time 
                  intervals due to flaring have been excluded.} 
\tablenotetext{b}{CCD chips that contain the specific targets in
                  the ACIS observations. The ellipses indicate  
                  observations not used.}
\tablenotetext{c}{A point source detected in NGC~5462. See Table \ref{tab2} 
                  for details.}
\end{deluxetable*}

The generation of X-ray-emitting hot gas reflects the star 
formation activities and the accompanied high energy processes. We have 
chosen to study the GHRs NGC~5461, NGC~5462, and NGC~5471 in M101 at a 
distance of 6.8~Mpc \citep[][ hence $1\arcsec$ = 33~pc]{2006ApJS..165..108S}, 
because they are several times as luminous as 30~Dor 
and exhibit contrasting optical morphologies \citep{2005ApJ...619..779C}, 
allowing us to investigate the generation and distribution of hot gas of 
GHRs with larger-scale starbursts. 

NGC~5461 is a large ($40\arcsec\times 25\arcsec$) H~II complex with multiple 
components \citep{1990ApJS...73..661H}. The linear size of NGC~5461 is 
$\sim 1.3$~kpc $\times$ 0.82~kpc. Six R136-class clusters of quite young 
ages ($<5$~Myr) have been identified by \citet{2005ApJ...619..779C}; five 
of them are located in H1105, the main body of NGC~5461. It is suggested 
that the entire region is associated with two generations of star formation 
events \citep{2006AJ....131..849P}. 

The GHR NGC~5462 corresponds to another large H~II complex spreading 
$48\arcsec\times33\arcsec$, or $\sim 1.6$~kpc~$\times$~1.1~kpc, as seen in ground-based 
H$\alpha$ images \citep{1975A&A....40..421I}.  Thirty-three H~II regions 
are identified within NGC~5462 \citep{1990ApJS...73..661H}, but none 
of them is comparable to 30~Dor ($\sim{}290$~pc$\times290$~pc).
\citet{2005ApJ...619..779C} have identified 25 cluster candidates 
in NGC~5462 and made photometric measurements; the two most luminous clusters 
have masses of $\sim{}2\times10^4\,M_\odot$ and ages of $\gtrsim 10$~Myr. 
At such ages, SNe explosions of massive stars begin 
to dominate the energetics of the GHRs \citep{2005MNRAS.361..679O}. 
If the resulting hot gas is confined therein, the GHRs may be 
expected to be luminous X-ray sources. 

NGC~5471 is a GHR in the outskirt of M101. \citet{1985ApJ...290..449S} 
identified five bright knots in NGC~5471 and concluded that the 
B-knot (NGC~5471B) contains a supernova remnant (SNR) because of its 
non-thermal radio emission and high [\ion{S}{2}]/H$\alpha$ ratio. 
This SNR was subsequently suggested to be a ``hypernova remnant'' because 
its $\sim{}10^{52}$~ergs explosion energy derived from the \emph{ROSAT} 
X-ray observation was much higher than the canonical SN explosion 
energy 10$^{51}$~ergs \citep{1999ApJ...517L..27W}. NGC~5471B has been observed 
to be a large-velocity-width source \citep[LVWS, ][]{1986ApJ...311...85C}, 
which is unique among NGC~5461, NGC~5462, and NGC~5471 
\citep{1994AJ....107..651Y}. In a follow-up critical examination of 
NGC~5471B, \citet{2002AJ....123.2462C} analyzed the underlying stellar
population and the spectral and kinematic properties of the nebula,
and concluded that these optical observations support the presence of a
hypernova remnant in NGC~5471B.

To carry out a comprehensive investigation of the physical properties of 
multi-phase gas and evolving processes in these GHRs, we have examined
\emph{Chandra} X-ray observations and obtained H$\alpha$ echelle spectral 
mapping of NGC~5461 and NGC~5471 with the 4m Mayall Telescope at Kitt 
Peak National Observatory (KPNO).  We report the analysis of hot gas 
in NGC~5461, NGC~5462, and NGC~5471 in this paper, and the
kinematics of warm ionized gas in these GHRs in an upcoming paper.
This paper is organized as follows: observations and data reductions 
are described in Section \ref{sec2}, the results of our analysis are 
presented in Section \ref{sec3}, the nature of the GHRs is discussed 
in Section \ref{sec4}, and conclusions are given in Section \ref{sec5}.

\section{Observations and Data Reductions}\label{sec2}

\begin{deluxetable*}{cccccccc}[!htp]
\tabletypesize{\scriptsize}
\tablecaption{Properties of Detected Point Sources\label{tab2}}
   \tablewidth{0pt}
  \tablehead{
  \colhead{Source} &
  \colhead{CXOU Name} &
  \colhead{$\delta_x$ ($\arcsec$)} &
  \colhead{CR $({\rm~cts~ks}^{-1})$} &
  \colhead{HR} &
  \colhead{HR1} &
  \colhead{Flag} &
  \colhead{MI} \\
  \noalign{\smallskip}
  \colhead{(1)} &
  \colhead{(2)} &
  \colhead{(3)} &
  \colhead{(4)} &
  \colhead{(5)} &
  \colhead{(6)} &
  \colhead{(7)} &
  \colhead{(8)}
}
\startdata
 1 &  J140338.65+541849.9 &  0.6 &$     0.07  \pm   0.02$& --& --& B, S & --\\
 2 &  J140341.28+541904.0 &  0.4 &$     0.77  \pm   0.06$& $-0.69\pm0.07$ & $ 0.35\pm0.08$ & B, S, H & NGC5457-X50, P107\\
 3 &  J140342.56+541910.5 &  0.5 &$     0.12  \pm   0.03$& --& --& B, S & --\\
 4 &  J140351.90+542149.4 &  0.3 &$     0.47  \pm   0.06$& $-0.88\pm0.06$ & $ 0    .28\pm0.13$ & B, S & NGC5457-X130\\
 5 &  J140352.60+542210.8 &  0.4 &$     0.12  \pm   0.03$& --& --& B, S & --\\
 6 &  J140353.83+542157.2 &  0.3 &$     5.24  \pm   0.18$& $-0.48\pm0.04$ & $ 0    .38\pm0.04$ & B, S, H & NGC5457-X21, P110\\
 7 &  J140354.30+542209.4 &  0.4 &$     0.13  \pm   0.03$& $-0.89\pm0.13$ & --& B, S & NGC5457-X280\\
 8 &  J140354.51+542152.0 &  0.5 &$     0.08  \pm   0.03$& --& --& B, H &  NGC5457-X243\\
 9 &  J140354.84+542135.3 &  0.4 &$     0.09  \pm   0.02$& --& --& B, S, H & --\\
10 &  J140429.21+542352.9 &  0.3 &$    13.40  \pm   1.60$& $-1.00\pm0.00$ & $-0.21\pm0.11$ & S, B & NGC5457-X17
\enddata
\tablecomments{The energy bands are defined as the following:
0.3--0.7 (S1), 0.7--1.5 (S2), 1.5--3 (H1), 3--7~keV (H2),
S = S1+S2, H = H1+H2, and B = S+H. 
Column (1): Source 
number. (2): {\sl Chandra} X-ray Observatory (unregistered) source name, 
following the {\sl Chandra} naming convention and the IAU Recommendation 
for Nomenclature (e.g., http://cdsweb.u-strasbg.fr/iau-spec.html). (3): 
Position uncertainty (1$\sigma$) calculated from the maximum likelihood 
centroiding and an approximate off-axis angle ($r$) dependent systematic 
error $0\farcs2+1\farcs4(r/8^\prime)^2$ (an approximation to Figure~4 of 
\citet{2002ApJ...574..258F}), which are added in quadrature.  
(4): On-axis source broad-band count rate --- the sum of the 
exposure-corrected count rates in the four bands. (5-6): The hardness 
ratios defined as 
${\rm HR}=({\rm H-S2})/({\rm H+S2})$ and ${\rm HR1}=({\rm S2-S1})/{\rm S}$,
listed only for values with uncertainties less than 0.2. (7): Energy
bands in which the source is detected, and from which the most accurate 
position is adopted in Column (2). (8): Matching identifiers of other X-ray
catalogs: from \citet[][ with a prefix "NGC5457-X"]{2011ApJS..192...10L} and
from \citet[][with a prefix "P"]{2001ApJ...561..189P}.
}
\end{deluxetable*}

\begin{deluxetable*}{ccccccc}[!hbp]
\tabletypesize{\scriptsize}
\tablecaption{Journal of IR, optical and UV data\label{tab1}}
\tablehead{
\colhead{Object} & \colhead{Observation Date} & \colhead{Filter} & 
\colhead{$\lambda$ (\AA)\tablenotemark{a}} & \colhead{Bandwidth (\AA)\tablenotemark{a}} & 
\colhead{Band}   & \colhead{Exposure (s)\tablenotemark{b}} }
\startdata
M101     & 2004 Mar. 08 & Channel 1 & 3.6$\mu$m	  & 0.75$\mu$m (21\%) & dust \& PAH & 10.4\\
\hline 
\noalign{\smallskip}
NGC~5461 & 1999 Mar. 23 & F656N & 6562 & 22  &H$\alpha$ & 160(1), 600(2)\\
         & 1999 Mar. 24 & F547M & 5454 & 487 &Str{\"o}mgren \emph{y}& 600(2), 100(2),20(3)\\
	     & 1999 Jun. 17 & F547M & 5454 & 487 &Str{\"o}mgren \emph{y}& 500(2)\\
NGC~5462 & 2000 Feb. 01 & F656N & 6562 & 22  &H$\alpha$ & 160(1), 600(2)\\
         & 2000 Feb. 01 & F547M & 5454 & 487 &Str{\"o}mgren \emph{y} & 600(1), 100(2), 20(2)\\
NGC~5471 & 1997 Nov. 01 & F656N & 6562 & 22  &H$\alpha$ & 180(1), 600(2)\\
\hline
\noalign{\smallskip}
NGC~5457 & Co-added     & FUV   & 1550 & 300 & hot star & 1500 
\enddata
\tablenotetext{a}{The effective wavelength and bandwidth of the filters 
                  are taken from the IRAC Instrument Handbook 
		  \citep{2004ApJS..154...10F} for the \emph{Spitzer} data, 
                  the WFPC2 Instrument Handbook \citep{1996wfpc.rept....5B}
		  for the \emph{HST} observations and the flight calibration 
                  \citep{2004AAS...205.2509M} for the \emph{GALEX} images.}
\tablenotetext{b}{The exposure time is followed by the number of exposures in
                  parentheses.  Multiple exposures are used to remove cosmic rays.}
\end{deluxetable*}

The main data sets used in this study are from the archive of the 
\emph{Chandra X-ray Observatory} observations. Archival data from the 
\emph{Hubble Space Telescope} (\emph{HST}), \emph{Spitzer Space Telescope} 
and \emph{Galaxy Evolution Explorer} (\emph{GALEX}) are also employed here 
for multi-wavelength comparisons.

The X-ray data analyzed here consist of several segments from the 1~Ms 
\emph{Chandra} observation of M101 \citep{2010ApJS..188...46K} and a 
15~ks \emph{Chandra} ACIS observation aimed at NGC~5471 (ObsID: 2779, 
PI: Q.~D.~Wang). The Ms observation is used for imaging and spectral 
analyses of NGC~5461 and NGC~5462.  NGC~5471 is located at $\sim20'$ 
off the aim point of the Ms observation of M101, where the 90\% 
energy-encircled radius (EER) is greater than $15\arcsec$ for a point 
source\footnote{\emph{Chandra} Proposer's Observatory Guide ver 12.0, 
Figure~4.13, available at http://cxc.harvard.edu/proposer/POG.}; 
therefore, only the 15~ks observation directly aimed at NGC~5471 is 
used to construct color composites of multi-band images. The small 
number of X-ray counts in this short observation prevents us from 
producing a tri-color X-ray image of NGC~5471. Table~\ref{tab3} 
summarizes the observations used for imaging and spectral analysis.  
Most of the adopted 
observations were made within one year span, so there is no significant 
variations of the instrumental background\footnote{http://cxc.harvard.edu/cal/Links/Acis/acis/Cal\_prods/bkgrnd/current/}. 

The CIAO 4.2 software is used for the X-ray data calibration and spectrum 
extraction. We reprocess the Chandra data starting with the Level 1 event 
files following the pipeline on the official website\footnote{http://cxc.harvard.edu/ciao/guides/acis\_data.html}.
Point-like sources in three broad bands, 0.3$-$1.5~keV~(S), 
1.5$-$7.0~keV~(H), and 0.3$-$7.0~keV~(B), are identified from the event 
maps, which are produced by merging all the available observations,
following the procedure detailed in \citet{2004ApJ...612..159W} 
using a combination of three algorithms: wavelet, sliding box, and maximum 
likelihood centroid fitting. The estimation of the count rate of a source 
is based on the number of counts within the 90\% EER determined from the 
calibrated point-spread function (PSF) of the instrument 
\citep{2000SPIE.4012...17J}. Point-like sources detected in the
three GHRs are listed in Table~\ref{tab2}. 

\begin{figure*}[!htp]
\centering
\includegraphics[scale=0.28]{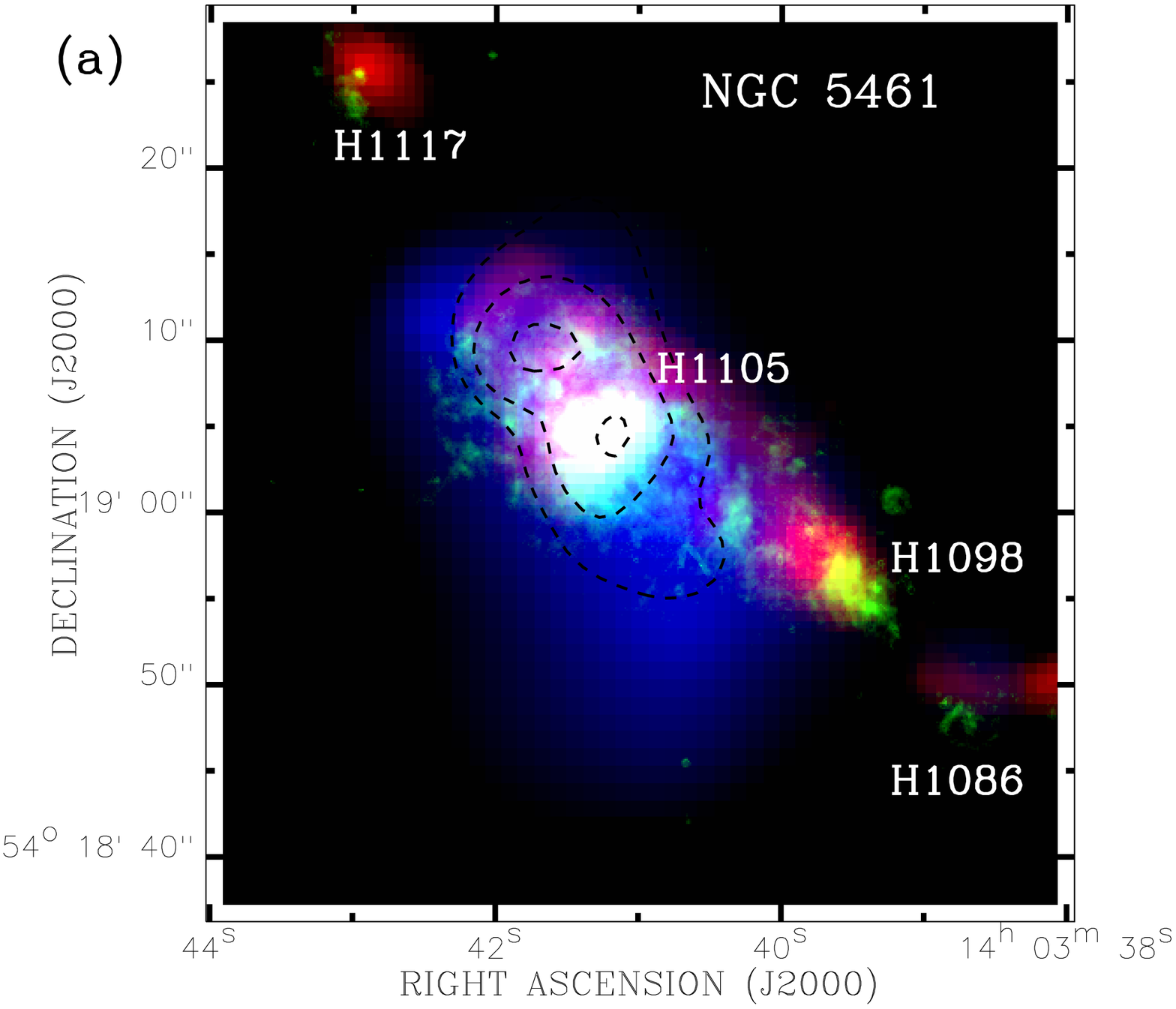}  
~~\includegraphics[scale=0.28]{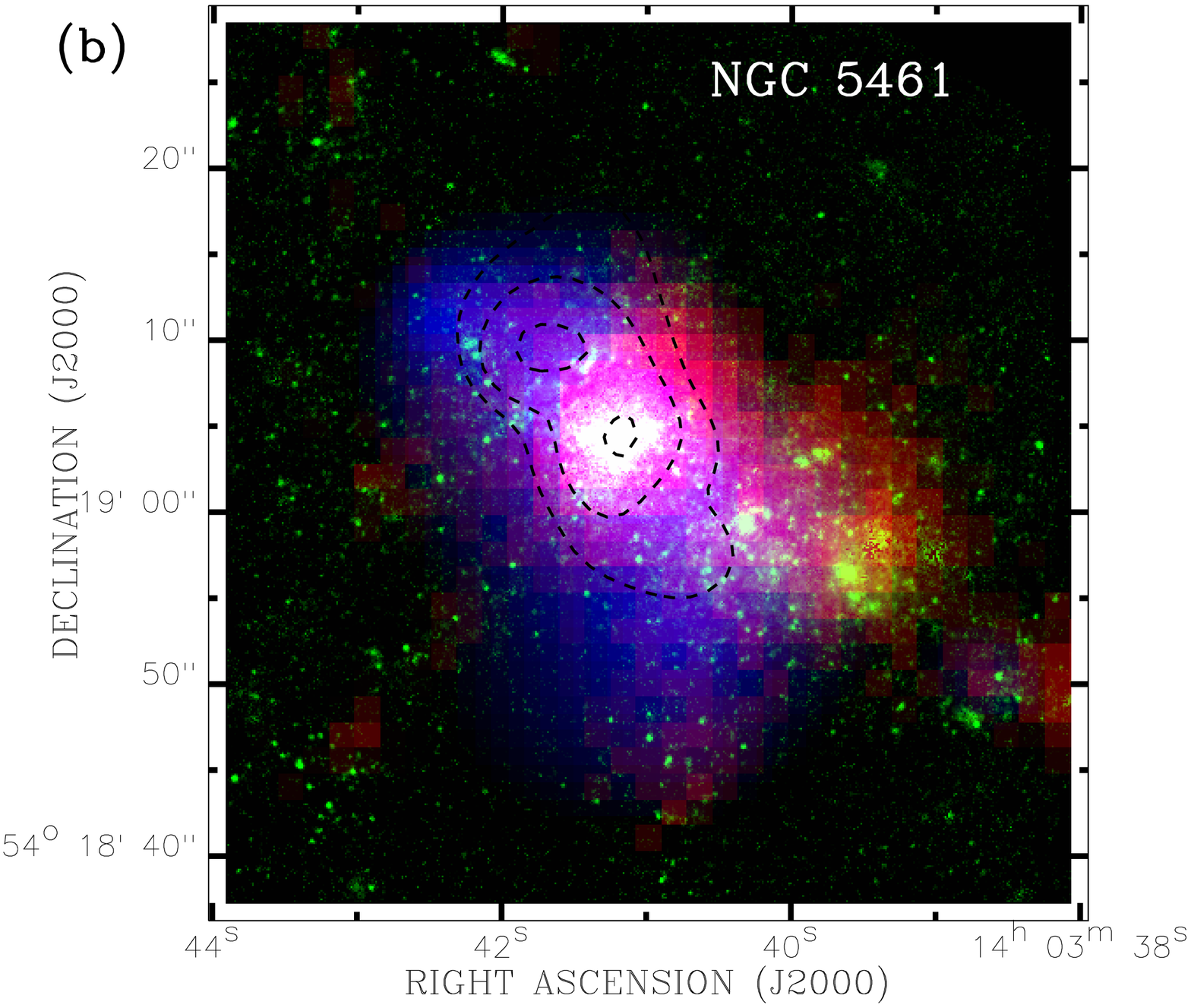}\\
\includegraphics[scale=0.28]{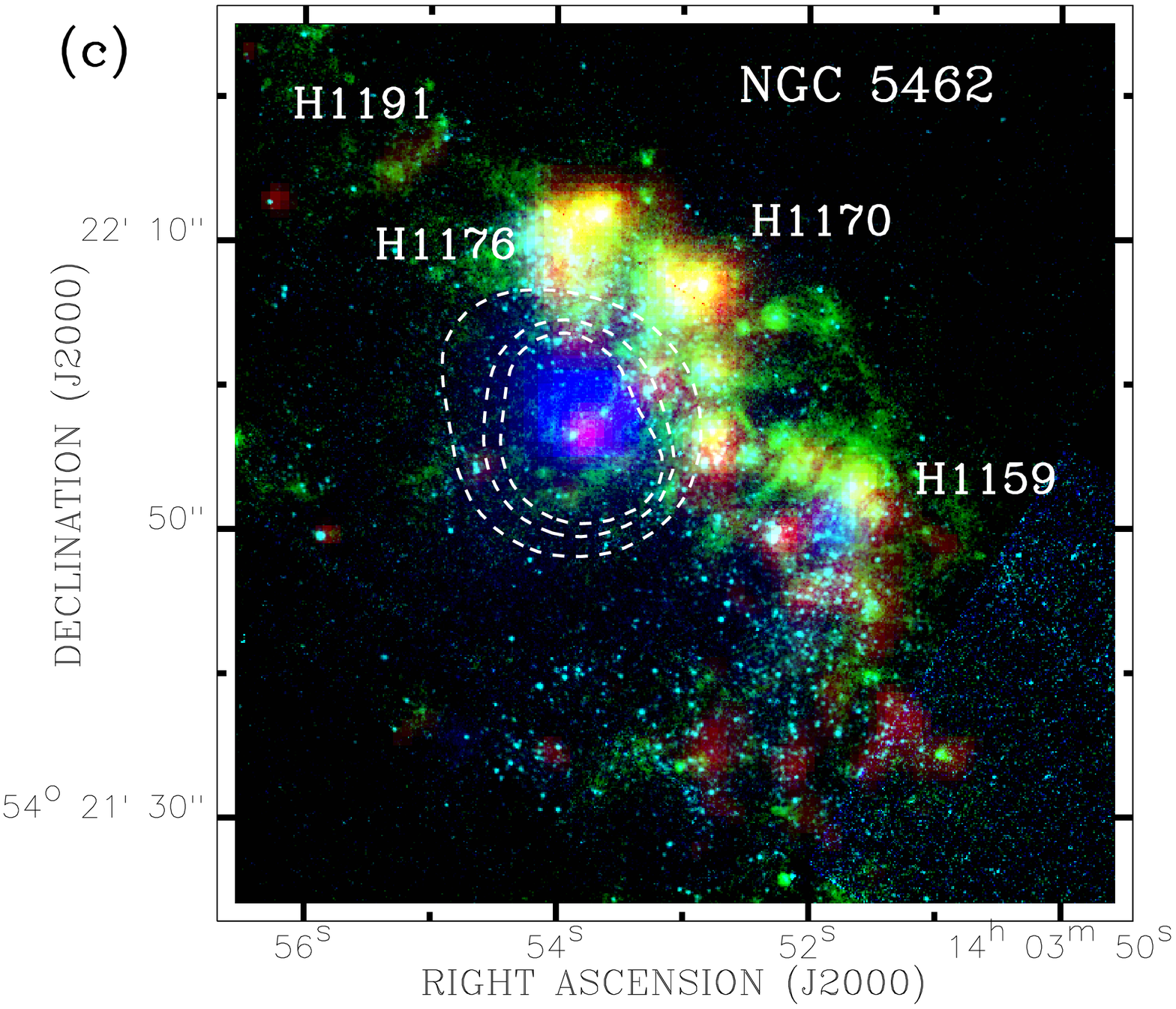}  
~~\includegraphics[scale=0.28]{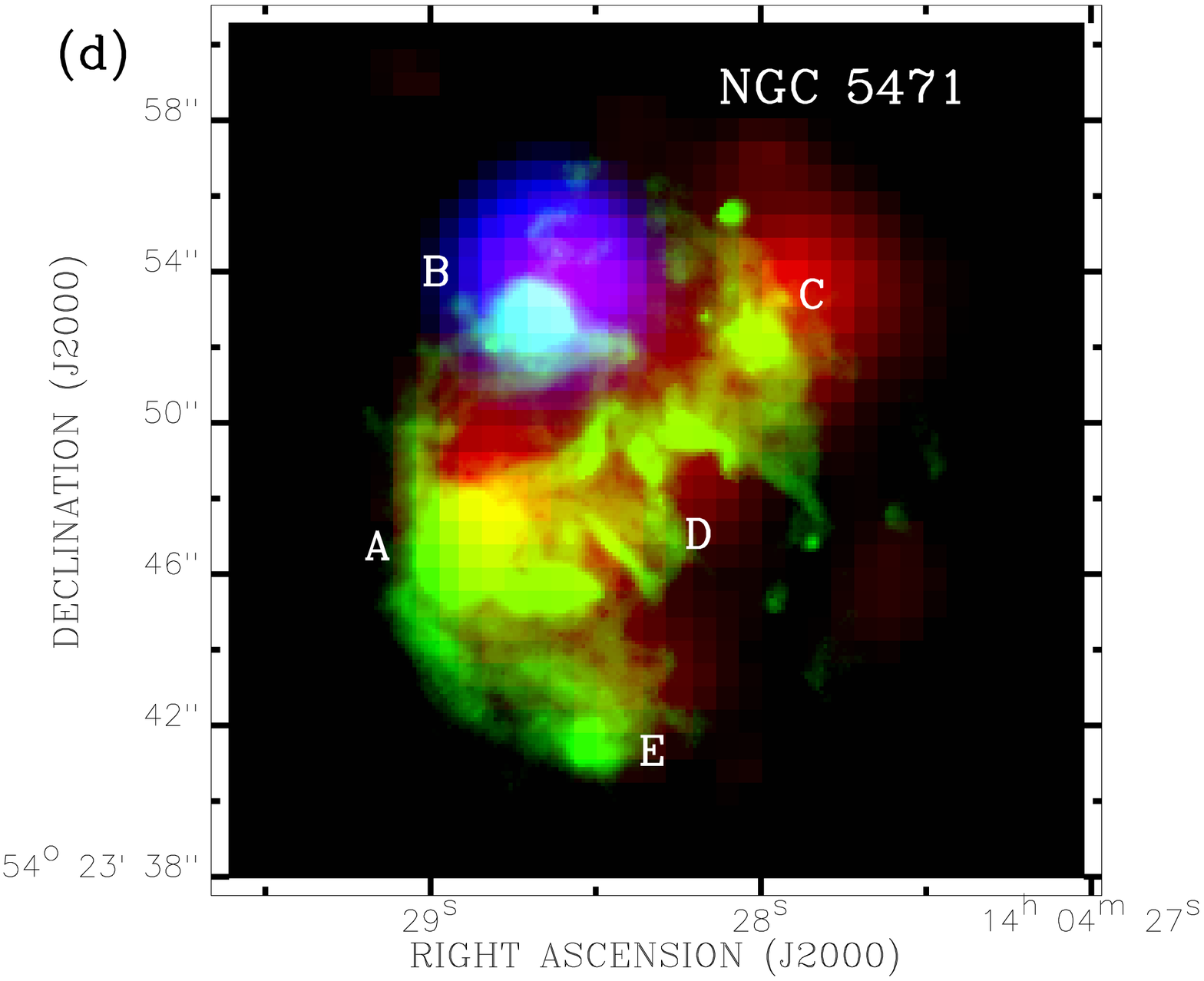}\\
\includegraphics[scale=0.28]{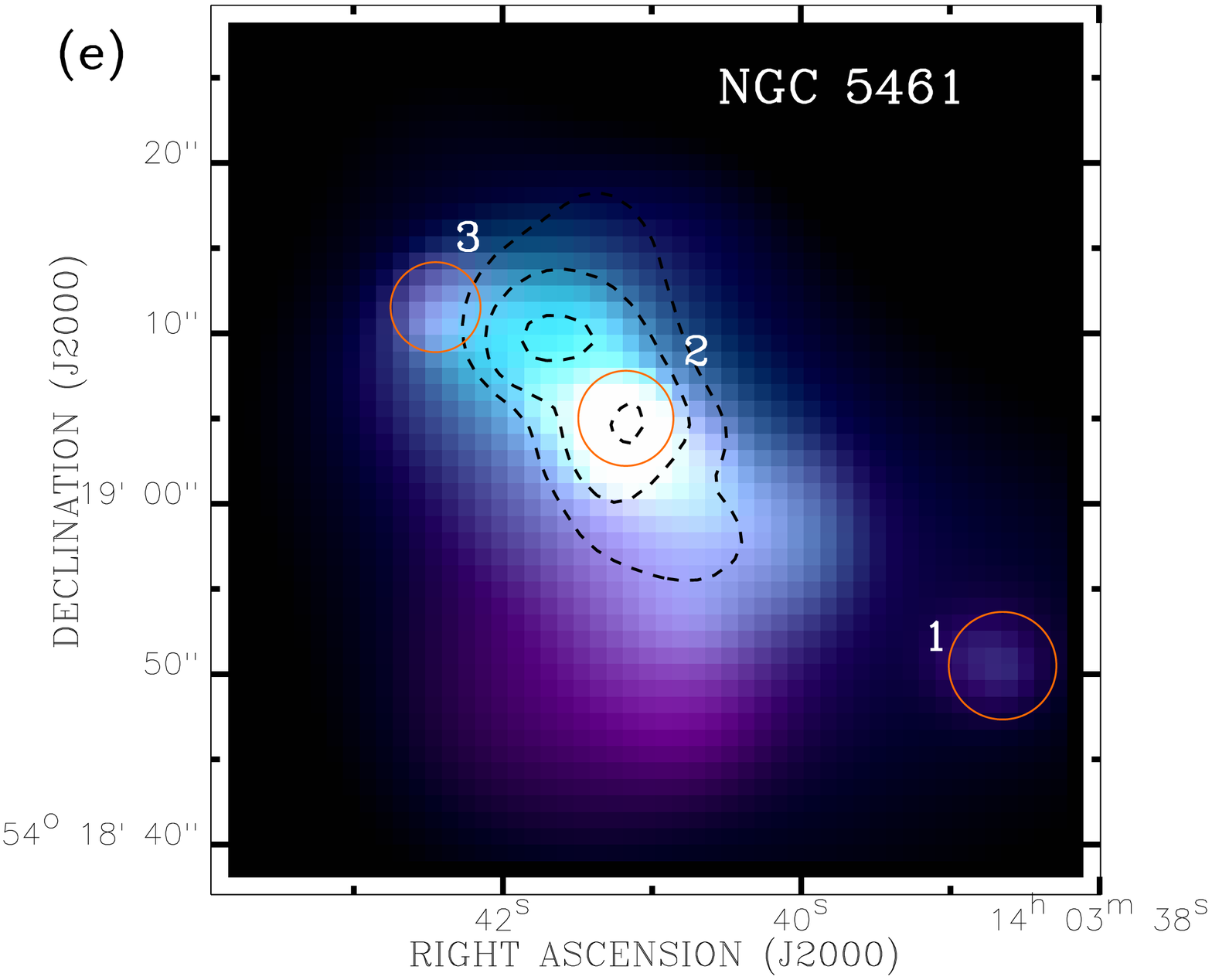}  
~~\includegraphics[scale=0.28]{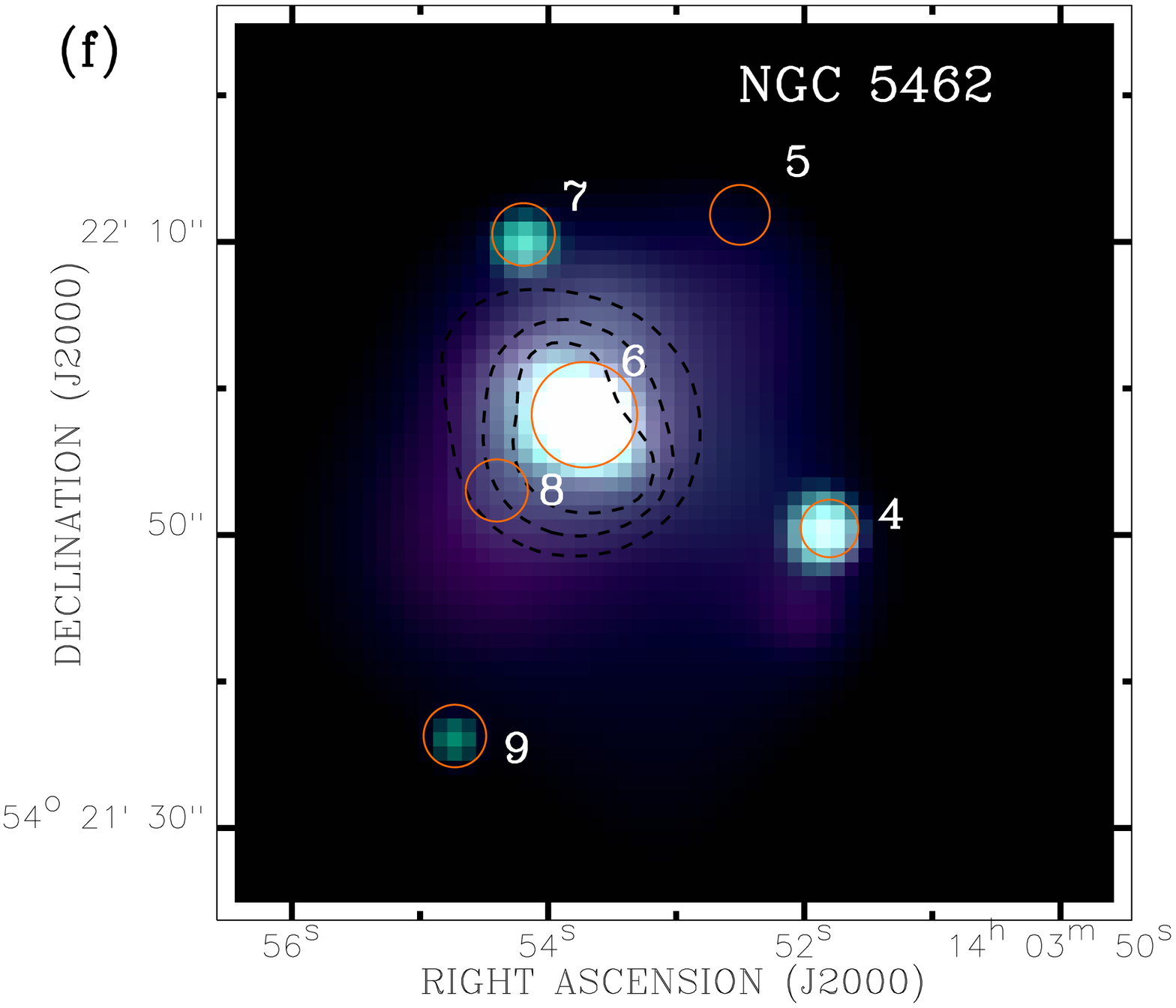}\\
\centering
\caption{Multi-band (panels {\em a--d}) and tri-color X-ray (panels 
         {\em e} and {\em f}) images of the three GHRs. 
         The red, green, and blue colors in the multi-band images {\em a}, 
         {\em c}, and {\em d} correspond to \emph{Spitzer} IRAC 3.6$\mu$m, 
         \emph{HST} WFPC2 F656N narrow band (H$\alpha$ line), and 
         \emph{Chandra} ACIS soft (0.3$-$1.5~keV) band;  those in 
	 panel~{\em b} represent the Str{\"o}mgren \emph{y} (\emph{HST} 
	 F547M), UV (\emph{GALEX}), and X-ray (\emph{Chandra} 0.3--1.5~keV) 
	 emission in NGC~5461; and those in the tri-color X-ray images 
	 {\em e} and {\em f} represent soft (0.3--0.7~keV), medium 
	 (0.7--1.5~keV) and hard (1.5--7.0~keV) bands, respectively, 
	 and the cyan color in panel~{\em c} represents the Str{\"o}mgren 
	 \emph{y} ({\em HST} F547M) band image of NGC~5462.
	 The orange circles in panels {\em e} and {\em f} mark the 
	 detected X-ray point sources; their radii are of 1.2 times the 
	 90\% ERR. The dashed contours overplotted on panels~{\em a, 
	 b, e, f}  (black), and {\em c} (white) represent the 
	 point-source-excised surface brightness of hard X-ray band 
         at levels of (1.19, 1.35, and 
	 1.68)$\times10^{-5}$~photon~s$^{-1}$~cm$^{-2}$~arcmin$^{-2}$
         in panels~{\em a, b}, and {\em e}, and (1.09, 1.29, and 
	 1.70)$\times10^{-5}$~photon~s$^{-1}$~cm$^{-2}$~arcmin$^{-2}$
	 in panels~{\em c} and {\em f}, respectively, which are 3$\sigma$, 
	 5$\sigma$, and 9$\sigma$ above the local background. \label{fig1}}
\end{figure*}

Different CCD chips have 
different responses, therefore, the instrumental backgrounds 
are subtracted from the raw event maps in four energy bands: 0.3$-$0.7~keV 
(S1), 0.7$-$1.5~keV (S2), 1.5$-$3.0~keV (H1), and 3.0$-$7.0~keV (H2), then 
the net count maps are corrected by the exposure maps of these four energy 
bands. The final net intensity maps (S1, S2, and H1+H2) are used to produce 
tri-color X-ray images, and the S1+S2 net intensity maps are used as the 
X-ray images in the multi-band color-composite images. Point sources 
have been excised from the H1+H2 intensity maps of NGC~5461 and NGC~5462, 
and the contours extracted 
from the resulting diffuse emission maps have been overplotted on the 
corresponding images in Figure~\ref{fig1}. 

The archival infrared (IR), optical and UV data from space telescopes are 
also included in this study in order to explore the physical properties 
of interstellar gas in various phases in and around these GHRs. 
The IR 3.6~$\mu$m image is obtained with the \emph{Spitzer} 
Infrared Array Camera (IRAC; PI: George Rieke). The \emph{HST} WFPC2
images of the three GHRs are from the Cycle 6 program GO-6829 
(PI: You-Hua Chu). The UV image is a stacked one from the ``Tracing 
the Extreme Edges of Galaxies in UV and HI'' program in Cycle 3 
(PI: Frank Bigiel). As we do not make photometric measurements, 
the post basic calibration data are used directly. Relevant 
information of these complementary IR, optical, and UV data is 
summarized in Table~\ref{tab1}. The multiple exposures of the 
\emph{HST} WFPC2 observations in each band are employed to remove cosmic 
rays using the IRAF tasks \emph{xzap} and 
\emph{imcombine}.

\section{Data Results}\label{sec3}

\subsection{Spatial Properties}

To investigate the relative distributions of the IR, optical, 
and X-ray emission, we have used these images to produce the color 
composites in Figure~\ref{fig1} (panels~{\em a}--{\em d}) for 
NGC~5461, NGC~5462, and NGC~5471. 
The X-ray images used in these color composites are the adaptively 
smoothed net intensity maps in the 0.3--1.5~keV energy band.
The alignment among the multi-band images is based on the world 
coordinates systems (WCS) and is accurate to better than $1\arcsec$.
To examine the hardness of the X-ray emission, we have also made
tri-color composites using the adaptively smoothed net intensity
maps in the 0.3--0.7, 0.7--1.5, and 1.5--7.0~keV bands for
NGC~5461 and NGC~5462, shown in the panels~{\em e} and {\em f} of 
Figure~\ref{fig1}. NGC~5471 has insufficient signal to warrant such 
color image.

\begin{figure*}[!htp]
\epsscale{0.8}
\centering
\includegraphics[scale=0.32]{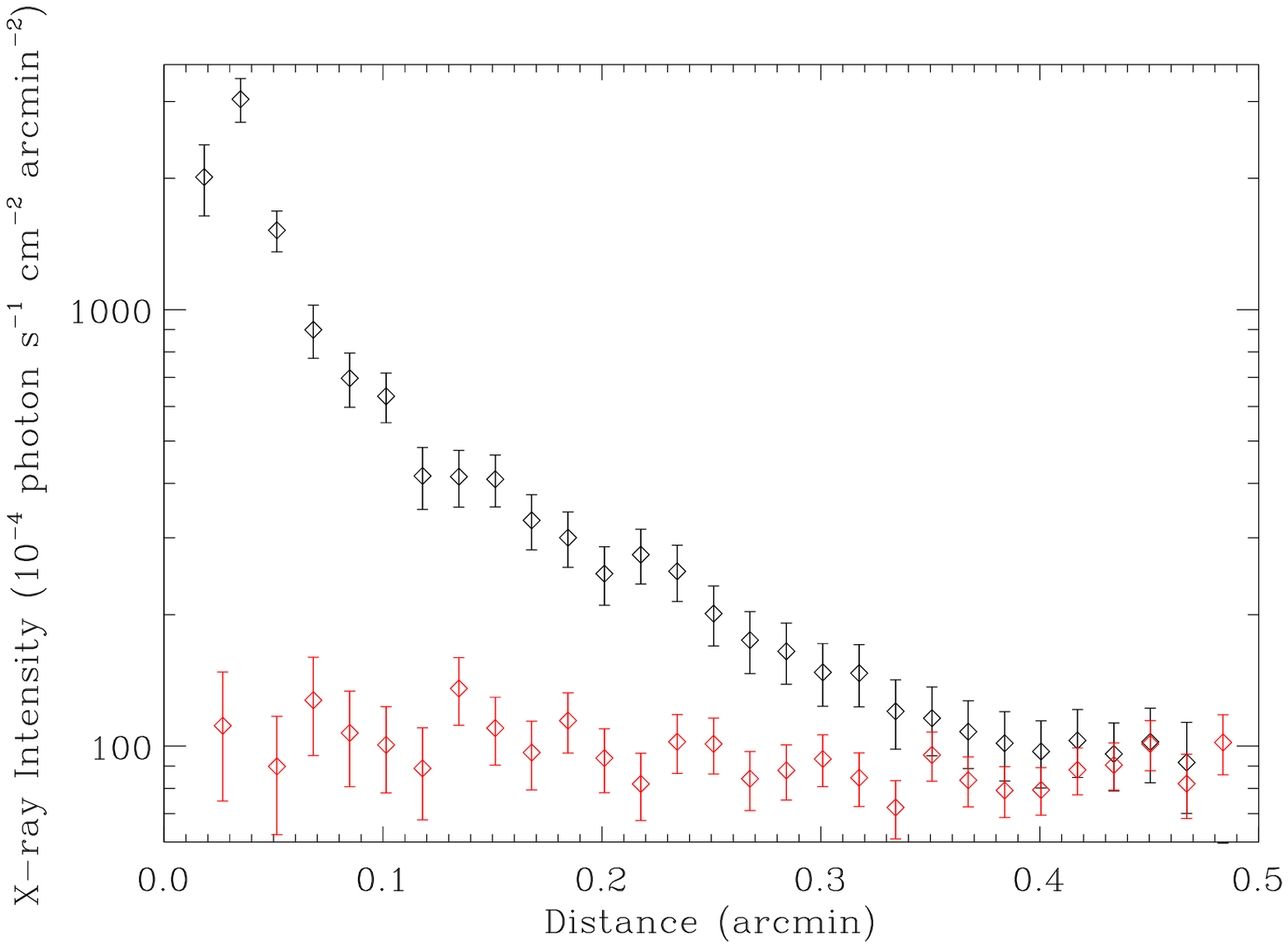}
\includegraphics[scale=0.32]{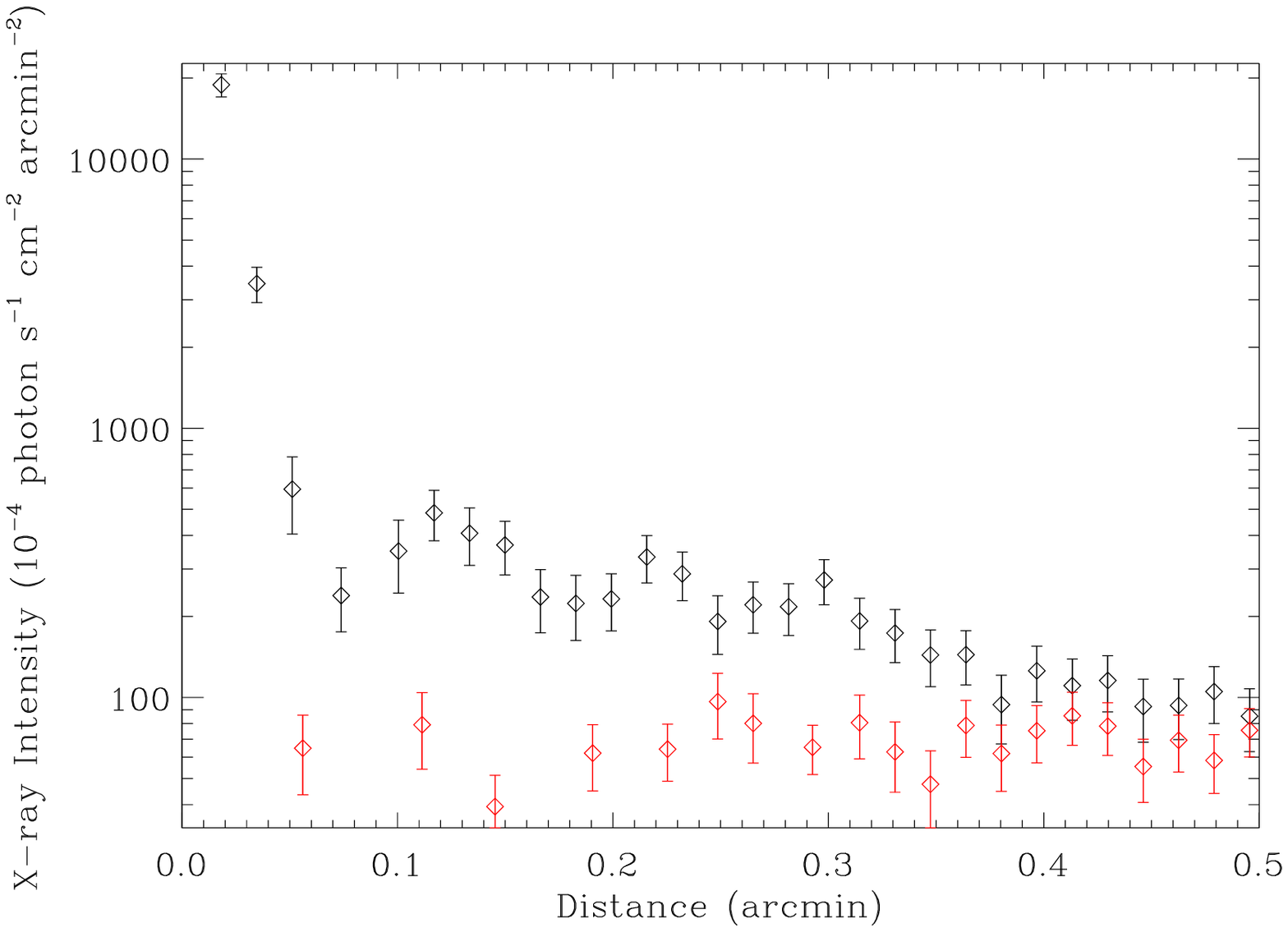}
\includegraphics[scale=0.32]{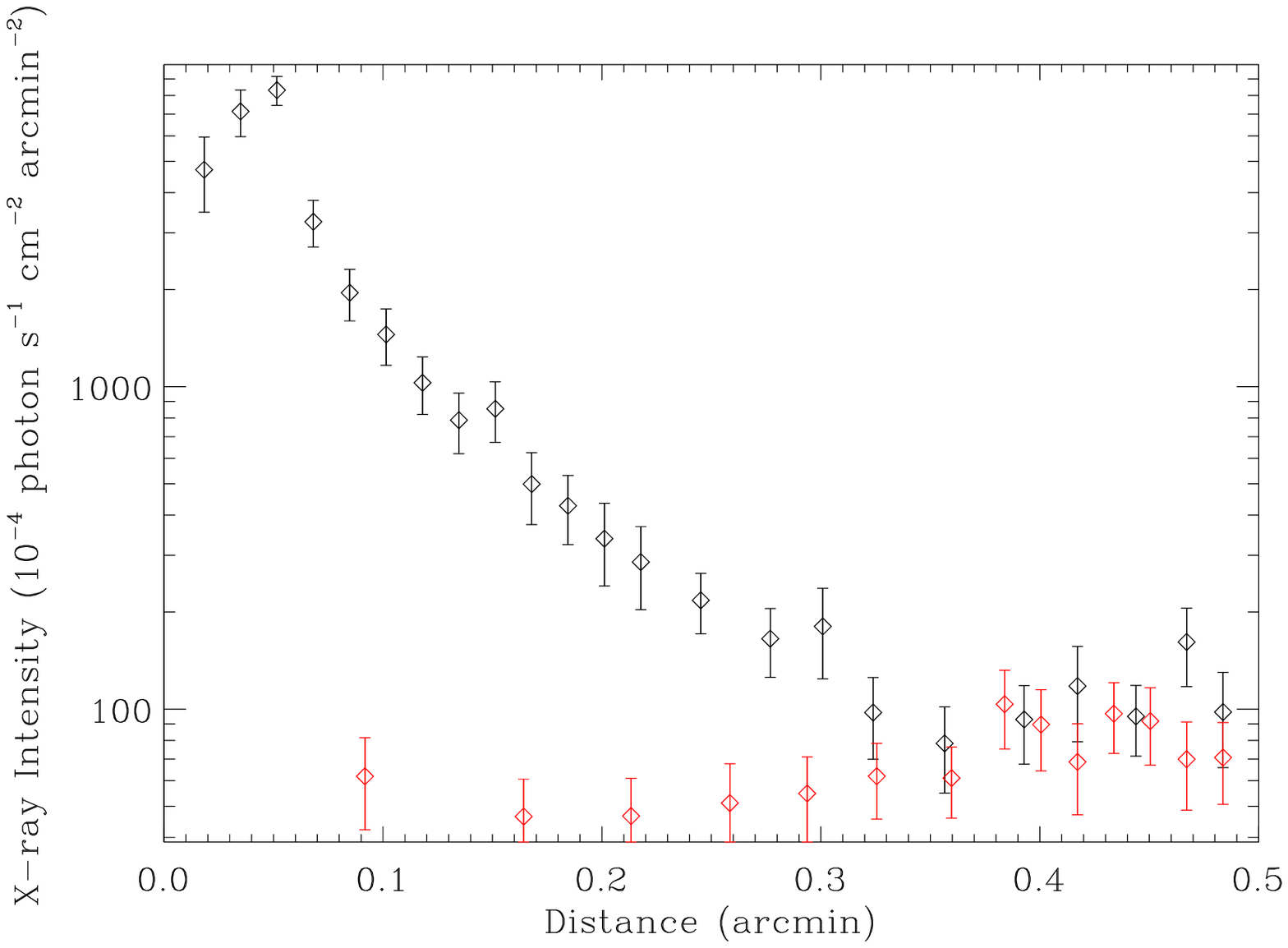}
\caption{Radial surface brightness profile comparisons in the broad band
         (0.3--7.0~keV) to determine the diffuse extraction regions in 
	 NGC~5461 (left), NGC~5462 (middle), and NGC~5471 (right). The 
	 distance to the center in NGC~5461 is measured along the minor
	 axes, and on- and off-source profiles are plotted in black and 
	 red data points, respectively. The extraction regions are 
	 selected to be elliptical with semi-major and minor axes of 
         26\farcs25$\times$21\arcsec\ and position angle of $44^\circ$ 
         for NGC~5461, circular with a radius of 27\arcsec\ for NGC~5462,
         and 24\farcs6\ for NGC~5471. The threshold count rate is listed 
         in the first row of Table~\ref{tab4} reporting spectral fitting 
         results. \label{fig2}}
\end{figure*}

\subsubsection{NGC~5461}
Optical images of NGC~5461 show a bright core, where a high concentration
of clusters are found, and extended distribution of stars and gas to the
northeast and southwest of the core, roughly along a spiral arm of M101
\citep[Figure~2 of ][]{2005ApJ...619..779C}. NGC~5461 is also a bright 
X-ray emitter (Figure~\ref{fig1}e), and as shown in Figure~\ref{fig1}a--b, its 
diffuse X-ray emission follows the distribution of stars, peaking at the 
bright optical core H1105 and extending to the northeast and southwest. 
The central X-ray emission has been identified as a quasi-soft source 
\citep[][ Src.\ 114 therein]{2004ApJ...609..710D}. In the 
0.3--0.7~keV band, the diffuse emission shows a remarkable deviation from 
the GHR's distribution -- it extends $\sim15''$ from the southern 
boundary of the H$\alpha$ bright region southward toward a region of 
elevated stellar density, which is visible in both \emph{HST} WFPC2 
continuum images \citep{2005ApJ...619..779C} and the \emph{GALEX} UV image 
(Figure~\ref{fig1}b). The linear scale of this soft X-ray extension is 
about 500~pc at the distance of M101, and is too large to be a sign of 
the blown out hot gas from H1105; the spectral analysis of
this feature is presented in Section \ref{x-ray_spec} and its nature
is further discussed in Section \ref{subsce}.

\subsubsection{NGC~5462}
Optical images of NGC~5462 show that its stars are distributed roughly 
in the northeast-southwest direction along a spiral arm, and that the
dense ionized gas (seen in H$\alpha$) is offset from the stars in the 
downstream direction \citep[northwest, ][]{2005ApJ...619..779C}.
The multi-band image of NGC~5462 in Figure~\ref{fig1}c shows 
that the diffuse X-ray emission is also offset from the stars, but in the 
opposite direction to the H~II gas. This relative distribution is very
different from that of NGC~5461, where diffuse X-rays follow the
stars and H~II gas except for the soft X-ray extension to the south.

As seen in the tri-color X-ray image (Figure~\ref{fig1}f), near the centroid 
of the diffuse X-ray emission is the bright X-ray point source Src.\ 6 
(as listed in Table~\ref{tab2}) at 
[14$^{\rm h}$03$^{\rm m}$53$^{\rm s}$.8, 54$^\circ21'57\arcsec$], which 
was identified as a black hole (BH) candidate by \citet{2003ApJ...582..184M}.
This point source is coincident with a bright IR source 
(Figure~\ref{fig1}c), and the candidate star cluster NGC~5462-19
of \citet{2005ApJ...619..779C}. Since this point source is the dominant 
X-ray emitter in this field, we analyze its spectrum in 
Section~\ref{x-ray_spec}. The non-thermal nature 
of its X-ray spectrum, the light absorption, and the steep photon 
index (see also Col.~4 in Table~\ref{tab4}) indicate that this point source 
could be a BH/X-ray binary in M101 or a nearly face-on 
AGN in the background. Src.\ 6 is very different from the point 
source Src.\ 4 near H1159, which is a quasi-soft X-ray source 
identified by \citet[][Src.\ 116 therein]{2004ApJ...609..710D}. The 
physical nature of Src.\ 6 will be discussed in Section~4.3.

\subsubsection{NGC~5471}
\emph{HST} images of NGC~5471 have resolved the five knots defined by 
\citet{1985ApJ...290..449S}, A--E, into clusters with associated
H~II gas \citep{2005ApJ...619..779C}. The multi-band image of NGC~5471 
(Figure~\ref{fig1}d) shows that the X-ray emission peaks at 
the B-knot. As this is a 15 ks observation, it is not clear whether faint 
diffuse emission is pervasive in this GHR complex. The angular size of 
NGC~5471B is comparable to the equivalent radius of the PSF even in the 
on-axis observation, and therefore it is not surprising that NGC~5471B 
is identified as a point source in our and former detections (Src.\ 10 
in Table~\ref{tab2}). The spectral analysis of this source uses all 
available sections of the Ms observation of M101 to improve the 
signal-to-noise ratio, assuming that all X-ray emission originates from 
the [S~II]/H$\alpha$ enhanced shell in NGC~5471B, as often seen in 
X-ray-bright superbubbles \citep[e.g., ][]{1990ApJ...365..510C}.

\begin{figure*}[!htp]
\plottwo{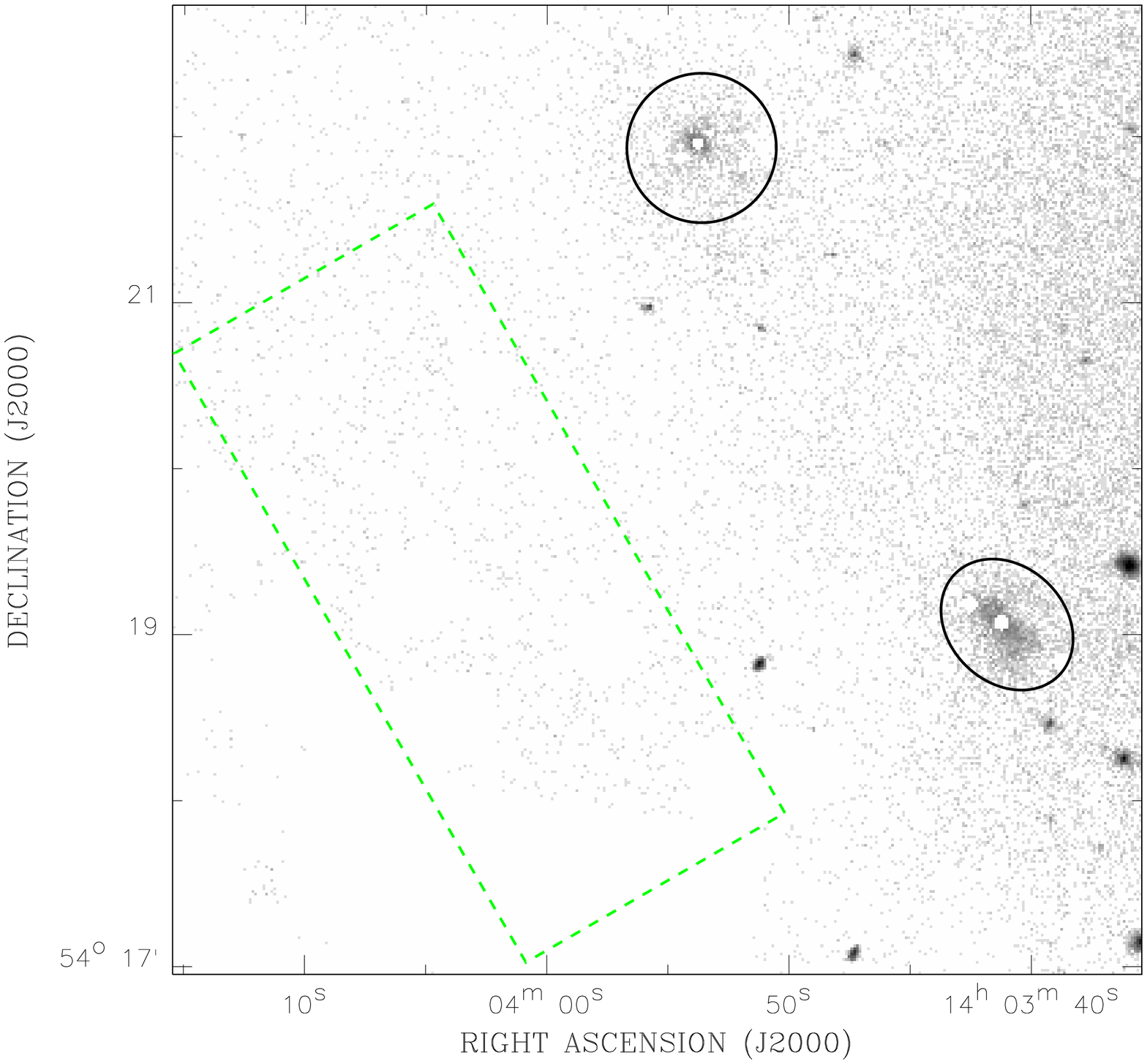}{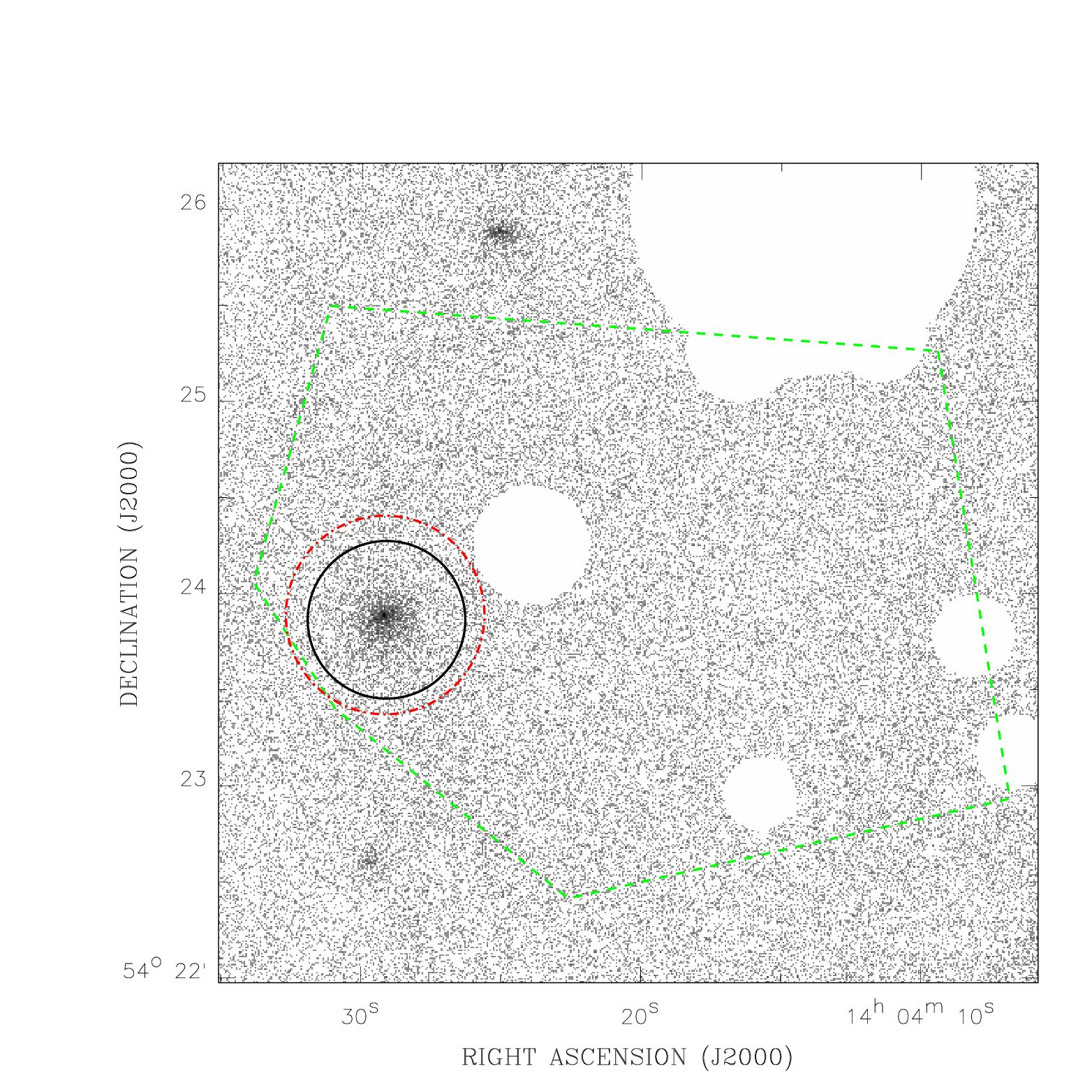}
\caption{Raw images of the fields containing NGC~5461, NGC~5462 (left) 
         and NGC~5471 (right). These are merged event maps of the ACIS 
         observations selected to make the spectral extraction of these 
         sources. On- and off-source extraction regions are represented 
         in black solid circles and green dashed polygons (outside the 
	 red dash-dotted circle for NGC~5471), respectively. The background 
	 regions are compromises among the data sets of different 
	 observations. \label{fig3}}
\end{figure*}

\subsection{X-ray Spectral Analysis}\label{x-ray_spec}
The X-ray spectra of the diffuse emission from the three GHRs are 
extracted from each individual observation using CIAO command 
\emph{specextract}. Six, seven, and eight observations are used 
for NGC~5461, NGC~5462, and NGC~5471 for full coverage, 
respectively (see Table~\ref{tab3} for the data adoption). The 
adopted on-source spectral extraction regions are elliptical 
(for NGC~5461) or circular (for NGC~5462 and NGC~5471). The region 
sizes are determined by comparing the on-source and off-source 
radial surface brightness profiles, as illustrated in 
Figure~\ref{fig2}.  The threshold intensities are listed in the 
first row of Table~\ref{tab4}, and the regions for the spectral 
extraction are overplotted on the raw X-ray images 
of NGC~5461, NGC~5462, and NGC~5471 in Figure~\ref{fig3}. All 
point-like sources detected in NGC~5461 and NGC~5462 are excised 
with circular exclusive regions of radii 1.2 times the 90\% EER, which 
enclose at least 95\% X-ray emission from the point sources. 
For the intriguing X-ray source Src.\ 6, 
11 observations are employed (Table~\ref{tab3}) to extract its 
spectrum using CIAO command \emph{psextract}. 
The adopted source region is circular with a radius 1.2 times the 90\% EER, 
and the background region is a concentric annulus with radii of 1.2--2 
times the 90\% EER.

Since the ACIS-S3 CCD is completely filled by M101 in every segment of 
the Ms observation, local backgrounds for the diffuse X-rays cannot be 
obtained from this chip,
and thus we adopt the ``double background subtraction'' method to account 
for the position dependence of the background, effective area, and energy 
response of the instruments. The off-source region for NGC~5461 or 
NGC~5462 is selected less ideally from an adjacent chip, and that 
for NGC~5471 is selected in the nearby field on the same chip (green dashed 
rectangle and polygon in Figure~\ref{fig3}). The spectra of the 
non-X-ray contributions are extracted individually from the corresponding 
regions in the stowed background data in CALDB~4.2 and normalized according 
to the 10-12~keV count rates \citep{2006ApJ...645...95H}. The on- and 
off-source spectra extracted from different observations for each GHR are 
weighted by exposure and combined to improve the statistics. 
The net on-source and off-source spectra are jointly 
fitted in XSPEC. 

\begin{figure*}[!htp]
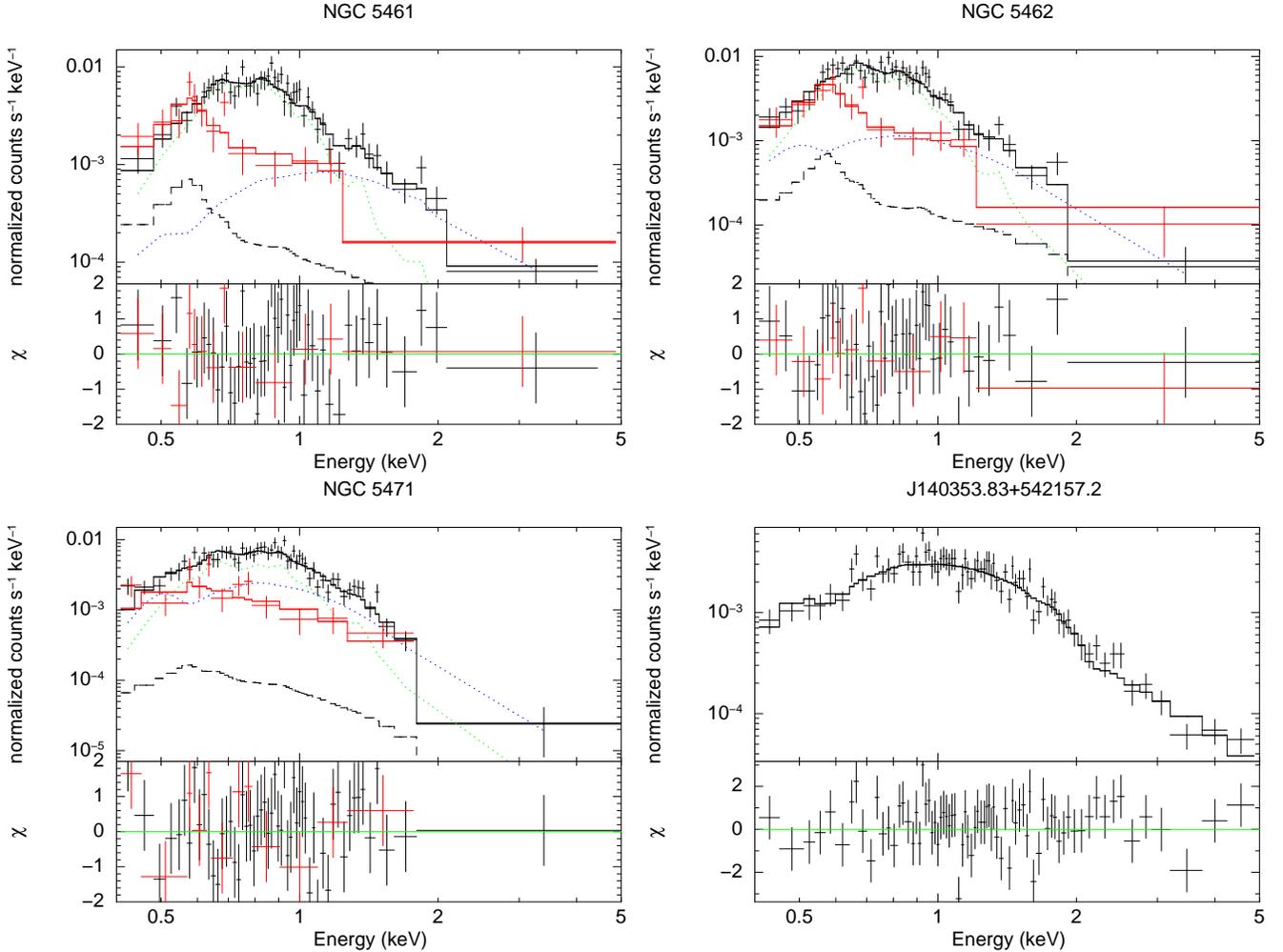

\centering
\epsscale{.80}
\includegraphics[scale=0.33,angle=-90]{figure4top-left.eps}
\includegraphics[scale=0.33,angle=-90]{figure4top-right.eps}\\
\includegraphics[scale=0.33,angle=-90]{figure4bottom-left.eps}
\includegraphics[scale=0.33,angle=-90]{figure4bottom-right.eps}
\caption{X-ray spectral analysis of NGC~5461 (top-left), NGC~5462 
         (top-right), NGC~5471 (bottom-left) and the point source 
         CXO~J140353.83+542157.2 (Src.\ 6) near NGC~5462 (bottom-right). 
         The on- and off-source spectra are plotted in black and red 
         colors, respectively. In the spectra of the GHRs, dotted lines 
	 represent thermal (green) and non-thermal (blue) contributions, 
	 and black dashed lines represent the diffuse background emissions
	 which are scaled from the off-source spectra. The 
	 best-fit model of Src.\ 6 is a 
         power-law, while the spectra of the three GHRs are well 
	 fitted with a {\tt MEKAL} (dotted green) plus power-law 
	 (dotted blue) model. 
         The spectral fit parameters are given in Table~\ref{tab4}.
         \label{fig4}}
\end{figure*}

\begin{deluxetable*}{lcccc}[!htp]
\tabletypesize{\footnotesize}
\tablecaption{X-ray spectroscopy of Emission from Giant H~II Regions \label{tab4}}
\tablewidth{0pt}
\tablehead{
\colhead{Emission Component} & \colhead{NGC~5461}  & \colhead{NGC~5462} & 
\colhead{Src.\ 6} & \colhead{NGC~5471B} 
}
\startdata
 (1) Threshold Count Rate\tablenotemark{a}(10$^{-6}$ sbu)     
     & $2.67\pm0.53$ & $16.3\pm 3.7$ & \nodata       & $2.14\pm0.56$ \\
 (2) Accumulated effective exposure (ks) & 228.9 & 238.5 & 381.3 & 276.6 \\  
 (3) Net Count Rate\tablenotemark{b} (10$^{-3}$ cts s$^{-1}$) 
     & $4.6\pm0.2$ & $4.3\pm0.2$ & $3.6\pm0.1$ & $4.5\pm0.1$ \\
 (4) Adopted Abundance for {\tt MEKAL} part (Z$_\odot$) 
     & 0.65  & 0.42              & \nodata  &  0.25         \\
 \hline 
 \noalign{\smallskip}
 (5) Neutral Hydrogen Column Density (10$^{22}$~cm$^{-2}$) 
     & $0.3^{+0.2}_{-0.2}$    & $0.2^{+0.2}_{-0.1}$ 
     & $0.26^{+0.05}_{-0.05}$    & $0.3^{+0.1}_{-0.2}$ \\
 (6) Temperature of {\tt MEKAL} Component (keV)   
     & $0.25^{+0.06}_{-0.06}$ & $0.24^{+0.04}_{-0.06}$ 
     & \nodata                & $0.19^{+0.08}_{-0.04}$ \\
 (7) Photon Index  
     & $2.4^{+0.9}_{-0.8}$    & $3.5^{+1.8}_{-1.1}$ 
     & $3.1^{+0.2}_{-0.2}$    & $5.6^{+1.0}_{-0.8}$ \\
 (8) Emission Measure/$d^2_{6.8}$ ($10^{62}\,{\rm cm}^{-3}$) 
     & $0.3^{+1.2}_{-0.2}$    & $0.2^{+1.7}_{-0.1}$   
     & \nodata                & $2.6^{+5.4}_{-2.1}$ \\
 (9) $A_\Gamma$ ($10^{-5}$photons~keV$^{-1}$~cm$^{-2}$~s$^{-1}$ at 1~keV) 
     & $0.3^{+0.2}_{-0.2}$    & $0.3^{+0.3}_{-0.1}$ 
     & $1.7^{+0.3}_{-0.2}$    & $1.4^{+0.6}_{-0.4}$ \\
 (10) $\chi^2$/degrees of freedom & 54.0/52 & 51.4/48 & 87.5/77 & 69.5/57 \\
 (11) $F_{\rm X,MEKAL}/F_{\rm X,PL}$\tablenotemark{c} 
     & 1.8  & 2.5             & \nodata  & 1.5 \\
 \hline 
 \noalign{\smallskip}
 (12) $L_{\rm X,0.5-5.0~keV}$\tablenotemark{d}($10^{38}$~erg~s$^{-1}$) 
     & 2.9   & 1.7            & 2.7      & 8.2 \\
 (13) $n_{\rm H}/f^{-1/2}d^{-1/2}_{6.8}$ ($\textrm{cm}^{-3}$)\tablenotemark{e} 
     & 0.29 &0.014 & \nodata  & 8.0 \\
 (14) $E_{\rm th}/f^{1/2}d^{5/2}_{6.8}$ ($10^{52}$~ergs)\tablenotemark{f}  
     & 9.1   & 161.           & \nodata  & 2.9 \\
\enddata
\tablecomments{See Section~\ref{x-ray_spec} and lower notes for the 
               descriptions of each row. The errors of the count rates are  
	       Poissonian, and the confidence ranges of the fitted parameters 
	       are at the 90\% level.}
\tablenotetext{a}{The count rates are counted in 0.3--7.0 keV energy 
                  interval without background subtraction, which are 
                  summarized from surface brightness profile 
		  analyses. 1~sbu=1~cts~cm$^{-2}$~s$^{-1}$~arcsec$^{-2}$.} 
\tablenotetext{b}{The net count rates are counted in 0.4--5.0~keV energy 
                  interval with instrumental background subtraction, 
                  and the emissions from point-like sources (Src.\ 1--9 in 
                  Table~\ref{tab2}) have been excluded from the GHRs' 
                  diffuse emissions.}
\tablenotetext{c}{Model absorbed flux ratio of {\tt MEKAL} to power-law 
                  of 0.5--5.0~keV.}
\tablenotetext{d}{Unabsorbed X-ray luminosity. Calculated from 
                  fitted model flux, for those three GHRs only the {\tt MEKAL} 
                  part is included.} 
\tablenotetext{e}{Root mean square (RMS) hydrogen density generated from 
                  the best-fit normalization parameter of the {\tt MEKAL} 
		  model, where $f$ denotes filling 
		  factor of the hot gas, the ellipsoids for H1105 (with 
		  half-axes $3\farcs15\times3\farcs15\times5\farcs3$, size 
		  of contour at 20\% of the top value) and NGC~5471B (with 
		  half-axes $1\farcs1\times0\farcs85\times0\farcs85$: size 
		  of the [S II]/H$\alpha$-enhanced shell), and the sphere 
		  for NGC~5462 (with radius $0\farcm45$) are assumed, and 
		  the He/H ratio are 0.09, 0.09, 0.08 for H1105, NGC~5462 
		  and NGC~5471B, respectively. }
\tablenotetext{f}{The thermal energy of the hot plasma is calculated as: 
                  $E_{\rm th}=V\cdot f\cdot u_{\rm X}=\frac{3}{2}\cdot 
                  2.3n_{\rm H}kT\cdot V\cdot f$, where $n_{\rm H}$ is the 
                  hydrogen density of shocked gas. }
\end{deluxetable*}

The optically-thin thermal plasma emission model, {\tt MEKAL} 
\citep{1985A&AS...62..197M}, plus an additional non-thermal component 
is used to fit the spectra of the diffuse emission from the three 
GHRs. Based on the oxygen and iron abundances determined from optical
spectrophotometry of the three GHRs \citep{1989ApJ...345..186T, 
2002ApJ...581..241E, 2003ApJ...591..801K}, the abundance parameters in the 
thermal components of NGC~5461, NGC~5462, and NGC~5471B are adopted to be 
0.65, 0.42, and 0.25~$Z_{\odot}$, respectively. The neutral hydrogen column 
density of the absorption to the fitting model is set to be free (while
it was fixed in the analyses of \citet{2010ApJS..188...46K} and 
\citet{1999ApJ...510L.139W}). The fitted spectra are shown 
in Figure~\ref{fig4}. The spectral results are summarized in 
Table~\ref{tab4}, in which Rows~(5)-(11) give the spectral fit 
parameters, and Rows~(12)-(14) show the derived physical 
parameters. 

The best-fit models for the X-ray spectra of the three GHRs are all 
characterized by a soft thermal component ($\sim{}0.2$~keV), and a 
non-negligible power-law component which dominates the spectra 
above $\sim1.0$~keV. The X-ray luminosities of the thermal component 
for NGC~5461, NGC~5462, and NGC~5471 are $(2.9, 1.7$, and 
$8.2) \times10^{38}$~erg~s$^{-1}$, respectively. As a comparison, 
\citet{2010ApJS..188...46K} and \citet{1999ApJ...510L.139W} obtained 
lower luminosities. Such discrepancies may be caused by a combination 
of a few factors: in addition to different spectral extraction regions,
larger absorbing hydrogen column densities and lower metal abundances 
were derived or adopted in our spectral fittings. It seems that these 
M101 GHRs are the most luminous in X-rays among 
the known GHRs in the local universe. This high X-ray luminosity will 
be discussed in Section~4.

We choose to use a power-law component to fit the hard tails 
in the spectra, instead of assuming another thermal component of higher 
temperature as \citet{2010ApJS..188...46K} did, because the power-law 
component may better represent the contribution from unresolved point 
sources, residual emission of the excluded point sources (in NGC~5462), 
and even emission from cosmic rays at shocks of stellar winds and SNRs. 

The point source CXOJ140353.83+542157.2 (Src.\ 6) is characterized by 
a power-law spectrum of intrinsic luminosity 
$\sim{}2.7\times10^{38}$~erg~s$^{-1}$ in 
the 0.5-5.0 energy band for the M101 distance of 6.8~Mpc. 
Given that the excluded region 
encircles $\gtrsim 90\%$ energy, the residual X-ray luminosity of the 
excluded point source in the diffuse emission of NGC~5462 may be up to 
$\sim10^{37}$~erg~s$^{-1}$. This source (Src.\ 6) is at least ten times 
brighter than the other detected point sources except for the central 
emissions in NGC~5461 and NGC~5471B, which are detected as two point 
sources (see Table~\ref{tab2}). 
Therefore, the contributions of point sources in the spectra of diffuse 
emission in the three GHRs may be at least $\sim10^{37}$~erg~s$^{-1}$, 
and are partly responsible for the power-law components in best-fit models.

The estimated hydrogen density of the hot gas, $n_{\rm H}$ (given in 
Table~\ref{tab4}), depends on its volume. The emission region of NGC~5461 
is irregular, and we assume that the elliptical extraction region represents 
a prolate ellipsoid and use the volume of the ellipsoid enclosed by the 
contour at 20\% of the peak surface brightness for the hot gas.
The hot gas volume in NGC~5462 is determined with the same assumptions.
The spectral extraction region of NGC~5471B (Figure~\ref{fig3}) is 
much larger than its optical size due to the large scattering of X-ray 
photons; thus, we assume that the diffuse X-ray emission originates from
the interior of the [S~II]/H$\alpha$-enhanced shell in NGC~5471B
\citep{2002AJ....123.2462C}, which 
is reasonable for a superbubble or a hypernova remnant scenario.
Consequently, we adopt the optical shell volume for the hot gas in
NGC~5471B. A volume filling factor $f$ is introduced to account for the 
fact that the emitting volume may be smaller than the assumed volume.

\section{Discussion}\label{sec4}

The X-ray luminosities of NGC~5461, NGC~5462, and NGC~5471 are at
least an order of magnitude higher than that of 30 Dor.  While 
the H$\alpha$ luminosities of these three M101 GHRs are also about
an order of magnitude higher than that of 30 Dor, the high X-ray
luminosities are not a direct consequence of the high H$\alpha$ 
luminosities.  As shown by \citet{2005ApJ...619..779C}, the three 
M101 luminous GHRs have multiple clusters at different evolutionary 
stages with ages ranging from $<$5 Myr to $>$10 Myr; furthermore, 
many clusters are as massive as the R136 cluster in 30 Dor.  The 
diffuse X-ray luminosity of hot gas generated by a cluster is 
time-dependent, and grows by almost two orders of magnitude at ages 
of 3-10 Myr when massive O stars explode as SNe \citep{2005MNRAS.361..679O}.  
Therefore, the high diffuse X-ray luminosities of the three M101 
GHRs are caused by the large number of massive clusters at ages when 
SN explosions are rampant.  Below, we discuss possible mechanisms 
for generating the hot gas in these GHRs.

\subsection{Superbubble Scenario of the GHRs}\label{subsce}

The X-ray emission associated with GHRs NGC~5461, NGC~5462, and 
NGC~5471 in M101 may be produced by similar processes as in superbubbles 
in the Magellanic Clouds \citep{1990ApJ...365..510C, 1991ApJ...373..497W, 
1991ApJ...374..475W}. Their soft appearances and luminosities are also 
as expected from the quasi-spherical superbubble model with a central 
stellar cluster with mass of $\sim 10^5 M_\odot$ \citep{1995ApJ...450..157C, 
2005ApJ...635.1116S}. However, the various spatial distributions of hot gas 
are not totally confined in visible superbubbles. 

At the southern outskirt of NGC 5461, where the soft X-ray emission extends, 
a bright patch of UV emission is present (see Figure~\ref{fig1}b), 
corresponding to the blue stars distributed in a triangular region that
is clearly seen in the color composite of the \emph{HST} WFPC2 images 
of \citet[Figure~2,][]{2005ApJ...619..779C}.
The main body of NGC~5461 stretches along a spiral arm, while the
afore-mentioned region of blue stars is detached from the main body and 
located in the upstream side, possibly belonging to an earlier episode 
of star formation triggered by the spiral density wave. The luminosity 
of the extended soft X-ray emission is of order $10^{37}$~erg~s$^{-1}$, 
if an absorbed thermal emission model with $N_{\rm H}\sim10^{21}$~cm$^{-2}$ 
and $kT\sim0.2$~keV is adopted. This X-ray luminosity is too high to be 
accounted for solely by the stellar emission from the unresolved OB-stars 
and active pre-main sequence stars that also contribute to the UV emission. 
This soft diffuse X-ray emission most likely originates from hot gas powered
by SN explosions and fast stellar winds.  
It is unlikely that outflows from the clusters in H1105 are
responsible for this hot gas because of the following reason. If the hot 
gas moves at its isothermal sound velocity $\sim$100 km~s$^{-1}$, it would
take 5 Myr to reach a distance of 500~pc, the outer bounary of the diffuse
X-ray emission.  As the clusters in H1105 are all $<$5 Myr old 
\citep{2005ApJ...619..779C}, it is 
impossible for them to have produced hot gas through SN explosions 5 Myr 
ago.  A better candidate for producing this hot gas is the star cloud, 
i.e., high surface density of stars over an extended area, between
H1105 and H1098, directly due north of the diffuse soft X-ray emission
and best seen in the color image of \citet{2005ApJ...619..779C}.  This star 
cloud is reminiscent of the stellar content in supergiant shells (SGSs),
such as SGSs LMC-3 and LMC-4, where diffuse soft X-ray emission is detected
with X-ray luminosities of 10$^{37}$ erg~s$^{-1}$ \citep{2001ApJS..136...99P}.
It is possible that the outflow of hot gas produced by energy feedback of this
star cloud contributes to the southern extension of diffuse soft X-ray emission.

In NGC~5462, the large displacement of the diffuse X-ray emission from the 
H~II regions indicates that the current young clusters embedded in the H~II 
regions are not likely responsible for generating the required 
X-ray-emitting hot gas.  On the other hand, the diffuse X-ray emission 
is better coincident with the ridge of older blue stars that are parallel 
to the H~II regions and offset in the same direction as the diffuse X-ray 
emission. It is possible that SN explosions from these stars power the 
X-ray emission.  The H$\alpha$ image of NGC~5462 
\citep[Figure 6e of ][]{2005ApJ...619..779C} shows long filaments extending 
from the ridge of blue stars toward the diffuse X-ray emission region, 
further supporting the hot gas outflow scenario.

It should be noted 
that quiescent superbubbles \citep[not energized by recent SN 
explosions;][]{1995ApJ...450..157C} associated with the sub-H~II regions 
in NGC~5462 would not be detected by even the 1~Ms \emph{Chandra} 
observation, whose detection limit for point sources is several 
$\times10^{36}$~erg~s$^{-1}$.  Superbubbles' soft X-ray emission can 
also be absorbed by surrounding dense cold medium and become undetectable. 
This may explain the apparent absence of X-ray emission associated with H1170. 
Only X-ray-bright superbubbles may be detected by the Ms \emph{Chandra} 
observation, and they appear as soft or quasi-soft sources, such as 
Src.\ 4 in H1159. 

The spatial analysis of X-rays in NGC~5471B is limited by the short exposure 
of the on-axis observation. The angular size of B-knot is too small to be 
resolved in the \emph{Chandra} observation, and the potential contribution 
from other knots of NGC~5471 to the total luminosity cannot be excluded. The 
analysis of high-dispersion echelle spectra of the H$\alpha$ line indicates
that only the B-knot is a LVWS source (FWHM$\gtrsim 120$~km~s$^{-1}$, 
Sun et~al., in preparation), which
is unique within NGC~5461 and NGC~5471. In the next section we discuss 
NGC~5471B as a hypernova remnant.

\subsection{NGC~5471B as Hypernova Remnant}\label{hnr}
The broad H$\alpha$ velocity profile of NGC~5471B suggests that an 
energetic explosive event has taken place. If the large velocities 
originate from an expanding shell, the extreme velocity offsets 
of the wings of the broad velocity profile indicate an expansion 
velocity at least 300~km~s$^{-1}$. 
If the temperature $T_{\rm X}$ measured from the {\tt MEKAL} model 
is adopted as the average temperature $\left<T\right>$ weighted by 
emission measure, which is 1.27 times the postshock temperature 
$T_{\rm s}$ based on the numerical calculations of the Sedov-Taylor 
self-similar solution, we can then estimate the blast wave velocity 
as $\sim340$ km s$^{-1}$, which is consistent with the value obtained 
by \citet{2002AJ....123.2462C}. We adopt the size of the remnant as 
that of the [\ion{S}{2}]/H$\alpha$-enhanced shell, and derive the 
atomic number density of ambient medium from the RMS hydrogen density 
(Row~12 in Table~\ref{tab4}) under the assumption that the filling 
factor is of unity. Based on the Sedov solution, we then estimate 
the remnant age as $3.4\times10^4$ yr, and the explosion energy as 
$1.5\times10^{52}$~ergs, comparable on order of magnitude 
to the thermal energy $E_{\rm th}$ (Table~\ref{tab4}) and consistent 
with the ``hypernova remnant'' scenario \citep{1999ApJ...517L..27W}. 

The remnant age is smaller than the typical time scale for it to enter the 
pressure-driven snowplow (PDS) phase \citep{1988ApJ...334..252C}: 
\begin{equation}
t_{\rm PDS}\sim 2.2\times10^4\,E^{3/14}_{52}n^{-4/7}_0\zeta^{-5/14}_{\rm m}\ {\rm yr},
\end{equation}
where $E_{52}$ is the expansion energy in units of 10$^{52}$~ergs, $n_0$ 
is the atomic number density of ambient medium, and $\zeta_{\rm m}$ is 
relative metallicity to solar abundance employed in the cooling function. 
In the case of NGC~5471B, $t_{\rm PDS}\simeq4.2\times10^4\,{\rm yr}$, 
which is consistent with the Sedov-Taylor phase assumption; 
meanwhile, the isothermal phase starts at radius: 
\begin{equation}
 R_{\rm PDS}=33.0\,E^{2/7}_{52}n^{-3/7}_0\zeta^{-1/7}_{\rm m,0.25}\ {\rm pc},
\end{equation}
which is comparable to the apparent size of NGC~5471B in the H$\alpha$ 
and [\ion{S}{2}] images.

For comparison, the quiescent superbubble model \citep{1995ApJ...450..157C} 
would produce unphysical parameters. In such a scenario, the temperature and 
density distributions in the shocked wind layer of an energy-conserving 
bubble are
\begin{equation}
 n(x)=n_{\rm c}(1-x)^{-2/5},\,T(x)=T_{\rm c}(1-x)^{2/5},
\end{equation}
where $x=r/R$ is the fractional radius, $R$ is the radius of the bubble:
\begin{equation}\label{equ1}
  R=42\,L_{37}^{1/5}n_0^{-1/5}t_{6}^{3/5}\;{\rm pc},
\end{equation}
$n_{\rm c}$ and $T_{\rm c}$ are the central density and temperature:
\begin{eqnarray}
 n_{\rm c} &=& 1.1\times10^{-2}\,L_{37}^{6/35}n_0^{19/35}t_{6}^{-22/35} 
               \;{\rm cm}^{-3},\\
 T_{\rm c} &=& 0.27\,L_{37}^{8/35}n_0^{2/35}t_{6}^{-6/35} 
               \;{\rm keV},
\end{eqnarray}
where $L_{37}$ is the mechanical luminosity of the stellar winds measured
in units of $10^{37}$~erg~s$^{-1}$, $n_0$ is the atomic number 
density of the ambient medium in units of cm$^{-3}$, and $t_6$ is the age 
of the bubble in $10^6$~yr.

Adopting the same definition of the dimensionless temperature $\tau=T_{\rm 
min}/T_{\rm c}$ as in \citet{1995ApJ...450..157C}, where $T_{\rm 
min}=5\times10^5$~K is the minimum temperature of the hot gas that will
emit in soft X-ray band, and assuming the measured X-rays represent the hot
gas from the interior of a spherical bubble with radius of 30~pc, the
average size of the [\ion{S}{2}]/H$\alpha$-enhanced shell, the volume emission 
measure of the hot gas in NGC~5471B can be expressed as
\begin{eqnarray}\label{equ2}
 EM &=& \int n_{\rm e}n_{\rm H} \ud^3 r\nonumber\\
    &=& 2.3\times10^{60}\,n_{\rm c}^2I(\tau)\;{\rm cm}^{-3},
\end{eqnarray}
where $n_{\rm c}$ is in units of cm$^{-3}$, and the dimensionless integral is
$I(\tau)=(125/33)-5\tau^{1/2}+(5/3)\tau^3-(5/11)\tau^{11/2}$; meanwhile, the
observed temperature is the average temperature:
\begin{equation}\label{equ3}
 \langle T\rangle = \frac{\int n(r)^2T(r)\ud^3r}{\int n(r)^2\ud^3r} 
                  = T_{\rm c}K(\tau)/I(\tau), 
\end{equation}
where $K(\tau)=(125/156)-(5/13)\tau^{13/2}+(5/4)\tau^4-(5/3)\tau^{3/2}$.
Therefore, substituting the radius of bubble $R$, the emission measure  and the
temperature of the hot gas to equations~\ref{equ1}, \ref{equ2}, and
\ref{equ3}, the derived mechanical luminosity of winds is
$1.3\times10^{38}$erg~s$^{-1}$, the atomic number density of ambient 
medium is several of $10^6$~cm$^{-3}$, and the age of the bubble is 41~Myr.

If the candidate clusters NGC~5471-10 and 11 are massive evolved ones,
$M_{\rm c}\sim{}10^4 M_\odot$, $t_{10}\gtrsim20$~Myr and
$t_{11}\gtrsim 60$~Myr, as seen in the color-magnitude diagram
\citep[][ Figure~10 therein]{2005ApJ...619..779C}, the mechanical luminosity and
the bubble age are consistent; however, the ISM density is not physical. 
Therefore, the ``hypernova remnant'' scenario is more favorable
compared to the superbubble model.

\subsection{Physical Properties of Src.\ 6 in NGC~5462}

The bright X-ray point source Src.\ 6 near the centroid 
of the diffuse emission in NGC~5462 has a spectrum that is 
best described by a power-law model.  Src.\ 6 has been proposed
to be a BH/X-ray binary \citep[P110 of ][]{2003ApJ...582..184M}. The 
coincidence between this BH/X-ray binary candidate and a 
previously identified optical cluster candidate 
\citep[NGC~5462-19 of ][]{2005ApJ...619..779C} appears to
suggest that the X-ray binary is a member of the cluster; 
however, a re-visit of the IR observations,
optical and radio images raises doubt about this picture.
 
\begin{figure}[!htp]
\epsscale{.80}
\centering
\includegraphics[scale=0.63]{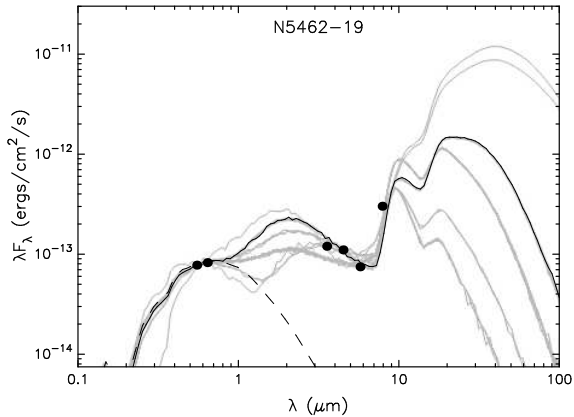}
\caption{Spectral energy distribution (SED) fitting of Src.\ 6 in the field 
         of NGC	5462. The 6 data points from the blue to red wings are of 
         Str{\"o}mgren~\emph{y} (\emph{HST}~F547M),
         WFPC2~R (\emph{HST}~F675W), \emph{Spitzer}~IRAC 3.6$\mu$m, 4.5$\mu$m,
         5.8$\mu$m, and 8.0$\mu$m bands, respectively. The dashed line
         represents the SED of a simple single population, which cannot
         fit the mid-IR excess; and the black and gray solid ones are
         scaled young stellar object (YSO) models which are not relevant here
         since this point source is too bright to be a YSO at the distance of
         M101. Therefore, this candidate cluster is more likely to be a
         background star-forming galaxy with a strong X-ray source if the
         detected sources all coincide.
         \label{fig5}}
\end{figure}

Bright IR emission is not expected from a BH/X-ray binary, yet 
\emph{Spitzer} observations detected a bright IR counterpart of 
Src.\ 6.  The spectral energy distribution (SED) of the optical 
and IR emission from this source (see Figure~\ref{fig5}) suggests 
the existence of heated dust.  This SED is inconsistent with that 
of young stellar objects (YSOs) in both spectral shapes and 
luminosities.  It is, however, similar to those frequently seen 
in active galactic nuclei (AGN).  If the minimum at the 5.8~$\mu$m 
band corresponds to the rest-frame minimum at the 4.5~$\mu$m band 
(due to the presence of PAH emission in the adjacent bands) seen 
in star-forming regions or galaxies, this source is likely a 
background AGN at $z \sim 0.3$.  The small absorption column implied 
by the best model fit to the X-ray spectrum can be reconciled if 
the AGN is viewed face-on.  

\begin{figure}[!htp]
\epsscale{.80}
\centering
\includegraphics[scale=0.33]{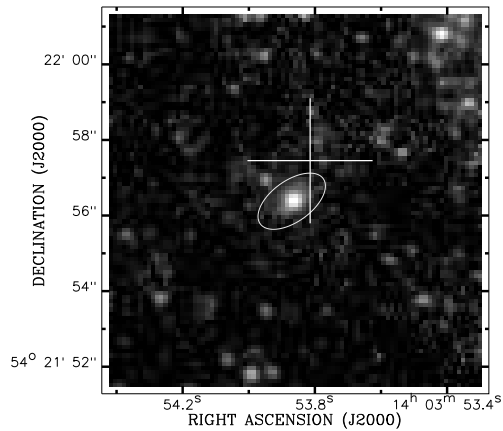}
\caption{Close-up view of the cluster candidate NGC~5462-19 
         \citep{2005ApJ...619..779C} in {\em HST} F547M band. The 
         white ellipse represents the exponential disk from the GALFIT 
         decomposition result with a radius of 5 times of the scale length, 
         and the white cross indicates the location of the X-ray detected 
         source (Src.\ 6). \label{fig6}}
\end{figure}

We examined the optical images of the cluster candidate \#19 more
closely.  The closeup $HST$ F547M image of NGC~5462-19 in 
Figure~\ref{fig6} shows a core and a diffuse halo, and it could 
be decomposited as a central point source and a exponential disk 
in GALFIT \citep{2010AJ....139.2097P}.  This morphology is 
consistent with that of an AGN in a disk galaxy.  If the
diffuse optical emission originates from a galactic disk, the 
scale length of the exponential disk, as 0\farcs2, implies a distance 
of several of 10$^9$ pc since the scale length is around 1 to 10~kpc 
for most disk galaxies. This estimated distance is of
the same order of magnitude as the redshift distance estimated
crudely from the IR SED.  For a distance of 10$^9$ pc, the intrinsic
X-ray luminosity of Src.\ 6 would be $\sim$10$^{43}$ erg~s$^{-1}$,
within the X-ray luminosity range of AGN \citep{2011ApJ...736...99A}. 
Another important clue to the nature of Src.\ 6 is provided by its
coincidence with the radio source NGC~5457E-$\theta$
\citep{2002ApJ...573..306E}, which definitively indicates that Src.\ 6
is a radio loud AGN rather than a BH/X-ray binary. We therefore
conclude that Src.\ 6 originates from the background
AGN, instead of a BH/X-ray binary in a candidate cluster.

\section{Concluding Comments}\label{sec5}

Giant H~II regions contain high concentrations of massive stars; therefore, 
they are excellent laboratories to study modes of massive star formation 
and subsequent evolution. We have analyzed \emph{Chandra} observations of 
X-ray emission from three GHRs in M101, NGC~5461, NGC~5471 and NGC~5462. 
The X-ray emission from these three GHRs exhibits diverse properties 
with respect to the individual H~II regions and associated stellar 
clusters. 
~~ \\

Our main results are the following:

1. The spectra of diffuse X-ray emission from NGC~5461, NGC~5462, and 
NGC~5471B all appear to be thermal with characteristic plasma temperatures 
$\sim$0.2~keV, although contributions from point sources may be significant 
and are modeled as an additional power-law component.  The luminosities of 
the thermal components of the GHRs are: $2.9\times10^{38}$, $1.7\times10^{38}$, 
and $8.2\times10^{38}$~erg~s$^{-1}$ for NGC~5461, NGC~5462, and NGC~5471B, 
respectively, which represent emission from hot gas commonly generated 
in massive star-forming regions.  They are very luminous because they contain
multiple massive clusters and star clouds at different ages and some of
them are ripe with rampant SN explosions. 

2. The distributions of diffuse X-ray emission in the three GHRs exhibit 
distinct features.  In NGC~5461, the brightest diffuse X-ray emission roughly
follows the distribution intense star formation; in addition, a patch of 
diffuse soft X-rays extend southward over $\sim$500 pc and may be attributed 
to outflows from the star cloud between H1105 and H1098.
In NGC~5462, the extended X-ray emission 
is displaced from the H~II regions and a ridge of blue stars; the
H$\alpha$ filaments extending from the ridge of star cloud to the
diffuse X-ray emission region suggests that hot gas outflows have 
occurred.  
In NGC~5471, the X-ray emission is very much confined within the B-knot, 
and our analysis indicates that the physical properties of the X-ray 
emitter are more consistent with a ``hypernova remnant'' scenario than 
a superbubble model. It is nevertheless unclear whether the explosion 
energy is provided by a single ``hypernova'' or concentrated 
multiple core-collapse SNe from a cluster.

3. The point source Src.\ 6 in NGC~5462 is most likely a background
AGN at a distance of order of 10$^9$ pc and its intrinsic X-ray luminosity
is $\sim$10$^{43}$ erg~s$^{-1}$.

\acknowledgments
We thank the anonmynous referee for her/his careful reading and insightful 
suggestions. YC acknowledges support from the 973 Program grant 
2009CB824800 and the NSFC grants 11233001 and 10725312. YHC acknowledges 
the support of SAO/CXC grant 
GO0-11025X. We also thank Bing Jiang for her assistance in data reduction, 
and Qiusheng Gu for helpful comments on the SED of NGC5462-19.

\bibliographystyle{apj}
\bibliography{reference,apj-jour}

\begin{thebibliography}{46}
\expandafter\ifx\csname natexlab\endcsname\relax\def\natexlab#1{#1}\fi

\bibitem[{{Allevato} {et~al.}(2011){Allevato}, {Finoguenov}, {Cappelluti},
  {Miyaji}, {Hasinger}, {Salvato}, {Brusa}, {Gilli}, {Zamorani}, {Shankar},
  {James}, {McCracken}, {Bongiorno}, {Merloni}, {Peacock}, {Silverman}, \&
  {Comastri}}]{2011ApJ...736...99A}
{Allevato}, V., {Finoguenov}, A., {Cappelluti}, N., {Miyaji}, T., {Hasinger},
  G., {et~al.} 2011, \apj, 736, 99

\bibitem[{{Biretta} {et~al.}(1996){Biretta}, {Ritchie}, {Baggett}, \&
  {MacKenty}}]{1996wfpc.rept....5B}
{Biretta}, J., {Ritchie}, C., {Baggett}, S., \& {MacKenty}, J. 1996,
  {Wavelength/Aperture Calibration of the WFPC2 Linear Ramp Filters}, Tech.
  rep.

\bibitem[{{Brandl}(2005)}]{2005ASSL..329...49B}
{Brandl}, B.~R. 2005, in Astrophysics and Space Science Library, Vol. 329,
  Starbursts: From 30 Doradus to Lyman Break Galaxies, ed. {R.~de Grijs \&
  R.~M.~Gonz{\'a}lez Delgado}, 49--+

\bibitem[{{Chen} {et~al.}(2002){Chen}, {Chu}, {Gruendl}, {Lai}, \&
  {Wang}}]{2002AJ....123.2462C}
{Chen}, C., {Chu}, Y., {Gruendl}, R., {Lai}, S., \& {Wang}, Q.~D. 2002, \aj,
  123, 2462

\bibitem[{{Chen} {et~al.}(2005){Chen}, {Chu}, \&
  {Johnson}}]{2005ApJ...619..779C}
{Chen}, C., {Chu}, Y., \& {Johnson}, K.~E. 2005, \apj, 619, 779

\bibitem[{{Chu} \& {Kennicutt}(1986)}]{1986ApJ...311...85C}
{Chu}, Y., \& {Kennicutt}, Jr., R.~C. 1986, \apj, 311, 85

\bibitem[{{Chu} {et~al.}(1995){Chu}, {Chang}, {Su}, \& {Mac
  Low}}]{1995ApJ...450..157C}
{Chu}, Y.-H., {Chang}, H.-W., {Su}, Y.-L., \& {Mac Low}, M.-M. 1995, \apj, 450,
  157

\bibitem[{{Chu} \& {Mac Low}(1990)}]{1990ApJ...365..510C}
{Chu}, Y.-H., \& {Mac Low}, M.-M. 1990, \apj, 365, 510

\bibitem[{{Cioffi} {et~al.}(1988){Cioffi}, {McKee}, \&
  {Bertschinger}}]{1988ApJ...334..252C}
{Cioffi}, D.~F., {McKee}, C.~F., \& {Bertschinger}, E. 1988, \apj, 334, 252

\bibitem[{{Di Stefano} \& {Kong}(2004)}]{2004ApJ...609..710D}
{Di Stefano}, R., \& {Kong}, A.~K.~H. 2004, \apj, 609, 710

\bibitem[{{Eck} {et~al.}(2002){Eck}, {Cowan}, \&
  {Branch}}]{2002ApJ...573..306E}
{Eck}, C.~R., {Cowan}, J.~J., \& {Branch}, D. 2002, \apj, 573, 306

\bibitem[{{Esteban} {et~al.}(2002){Esteban}, {Peimbert}, {Torres-Peimbert}, \&
  {Rodr{\'{\i}}guez}}]{2002ApJ...581..241E}
{Esteban}, C., {Peimbert}, M., {Torres-Peimbert}, S., \& {Rodr{\'{\i}}guez}, M.
  2002, \apj, 581, 241

\bibitem[{{Fazio} {et~al.}(2004){Fazio}, {Hora}, {Allen}, {Ashby}, {Barmby},
  {Deutsch}, {Huang}, {Kleiner}, {Marengo}, {Megeath}, {Melnick}, {Pahre},
  {Patten}, {Polizotti}, {Smith}, {Taylor}, {Wang}, {Willner}, {Hoffmann},
  {Pipher}, {Forrest}, {McMurty}, {McCreight}, {McKelvey}, {McMurray}, {Koch},
  {Moseley}, {Arendt}, {Mentzell}, {Marx}, {Losch}, {Mayman}, {Eichhorn},
  {Krebs}, {Jhabvala}, {Gezari}, {Fixsen}, {Flores}, {Shakoorzadeh}, {Jungo},
  {Hakun}, {Workman}, {Karpati}, {Kichak}, {Whitley}, {Mann}, {Tollestrup},
  {Eisenhardt}, {Stern}, {Gorjian}, {Bhattacharya}, {Carey}, {Nelson},
  {Glaccum}, {Lacy}, {Lowrance}, {Laine}, {Reach}, {Stauffer}, {Surace},
  {Wilson}, {Wright}, {Hoffman}, {Domingo}, \& {Cohen}}]{2004ApJS..154...10F}
{Fazio}, G.~G., {Hora}, J.~L., {Allen}, L.~E., {Ashby}, M.~L.~N., {Barmby}, P.,
  {et~al.} 2004, \apjs, 154, 10

\bibitem[{{Feast} {et~al.}(1960){Feast}, {Thackeray}, \&
  {Wesselink}}]{1960MNRAS.121..337F}
{Feast}, M.~W., {Thackeray}, A.~D., \& {Wesselink}, A.~J. 1960, \mnras, 121,
  337

\bibitem[{{Feigelson} {et~al.}(2002){Feigelson}, {Broos}, {Gaffney}, {Garmire},
  {Hillenbrand}, {Pravdo}, {Townsley}, \& {Tsuboi}}]{2002ApJ...574..258F}
{Feigelson}, E.~D., {Broos}, P., {Gaffney}, III, J.~A., {Garmire}, G.,
  {Hillenbrand}, L.~A., {Pravdo}, S.~H., {Townsley}, L., \& {Tsuboi}, Y. 2002,
  \apj, 574, 258

\bibitem[{{Hickox} \& {Markevitch}(2006)}]{2006ApJ...645...95H}
{Hickox}, R.~C., \& {Markevitch}, M. 2006, \apj, 645, 95

\bibitem[{{Hodge} {et~al.}(1990){Hodge}, {Gurwell}, {Goldader}, \&
  {Kennicutt}}]{1990ApJS...73..661H}
{Hodge}, P.~W., {Gurwell}, M., {Goldader}, J.~D., \& {Kennicutt}, Jr., R.~C.
  1990, \apjs, 73, 661

\bibitem[{{Hunter} {et~al.}(1996){Hunter}, {Baum}, {O'Neil}, \&
  {Lynds}}]{1996ApJ...456..174H}
{Hunter}, D.~A., {Baum}, W.~A., {O'Neil}, Jr., E.~J., \& {Lynds}, R. 1996,
  \apj, 456, 174

\bibitem[{{Israel} {et~al.}(1975){Israel}, {Goss}, \&
  {Allen}}]{1975A&A....40..421I}
{Israel}, F.~P., {Goss}, W.~M., \& {Allen}, R.~J. 1975, \aap, 40, 421

\bibitem[{{Jerius} {et~al.}(2000){Jerius}, {Donnelly}, {Tibbetts}, {Edgar},
  {Gaetz}, {Schwartz}, {Van Speybroeck}, \& {Zhao}}]{2000SPIE.4012...17J}
{Jerius}, D., {Donnelly}, R.~H., {Tibbetts}, M.~S., {Edgar}, R.~J., {Gaetz},
  T.~J., {Schwartz}, D.~A., {Van Speybroeck}, L.~P., \& {Zhao}, P. 2000, in
  Society of Photo-Optical Instrumentation Engineers (SPIE) Conference Series,
  Vol. 4012, Society of Photo-Optical Instrumentation Engineers (SPIE)
  Conference Series, ed. {J.~E.~Truemper \& B.~Aschenbach}, 17--27

\bibitem[{{Kennicutt}(1984)}]{1984ApJ...287..116K}
{Kennicutt}, Jr., R.~C. 1984, \apj, 287, 116

\bibitem[{{Kennicutt} {et~al.}(2003){Kennicutt}, {Bresolin}, \&
  {Garnett}}]{2003ApJ...591..801K}
{Kennicutt}, Jr., R.~C., {Bresolin}, F., \& {Garnett}, D.~R. 2003, \apj, 591,
  801

\bibitem[{{Kuntz} \& {Snowden}(2010)}]{2010ApJS..188...46K}
{Kuntz}, K.~D., \& {Snowden}, S.~L. 2010, \apjs, 188, 46

\bibitem[{{Liu}(2011)}]{2011ApJS..192...10L}
{Liu}, J. 2011, \apjs, 192, 10

\bibitem[{{Mewe} {et~al.}(1985){Mewe}, {Gronenschild}, \& {van den
  Oord}}]{1985A&AS...62..197M}
{Mewe}, R., {Gronenschild}, E.~H.~B.~M., \& {van den Oord}, G.~H.~J. 1985,
  \aaps, 62, 197

\bibitem[{{Morrissey} \& {GALEX Science}(2004)}]{2004AAS...205.2509M}
{Morrissey}, P., \& {GALEX Science}. 2004, in Bulletin of the American
  Astronomical Society, Vol.~36, American Astronomical Society Meeting
  Abstracts, 1385--+

\bibitem[{{Mukai} {et~al.}(2003){Mukai}, {Pence}, {Snowden}, \&
  {Kuntz}}]{2003ApJ...582..184M}
{Mukai}, K., {Pence}, W.~D., {Snowden}, S.~L., \& {Kuntz}, K.~D. 2003, \apj,
  582, 184

\bibitem[{{Oskinova}(2005)}]{2005MNRAS.361..679O}
{Oskinova}, L.~M. 2005, \mnras, 361, 679

\bibitem[{{Pellerin}(2006)}]{2006AJ....131..849P}
{Pellerin}, A. 2006, \aj, 131, 849

\bibitem[{{Pence} {et~al.}(2001){Pence}, {Snowden}, {Mukai}, \&
  {Kuntz}}]{2001ApJ...561..189P}
{Pence}, W.~D., {Snowden}, S.~L., {Mukai}, K., \& {Kuntz}, K.~D. 2001, \apj,
  561, 189

\bibitem[{{Peng} {et~al.}(2010){Peng}, {Ho}, {Impey}, \&
  {Rix}}]{2010AJ....139.2097P}
{Peng}, C.~Y., {Ho}, L.~C., {Impey}, C.~D., \& {Rix}, H.-W. 2010, \aj, 139,
  2097

\bibitem[{{Points} {et~al.}(2001){Points}, {Chu}, {Snowden}, \&
  {Smith}}]{2001ApJS..136...99P}
{Points}, S.~D., {Chu}, Y.-H., {Snowden}, S.~L., \& {Smith}, R.~C. 2001, \apjs,
  136, 99

\bibitem[{{Saha} {et~al.}(2006){Saha}, {Thim}, {Tammann}, {Reindl}, \&
  {Sandage}}]{2006ApJS..165..108S}
{Saha}, A., {Thim}, F., {Tammann}, G.~A., {Reindl}, B., \& {Sandage}, A. 2006,
  \apjs, 165, 108

\bibitem[{{Schaerer} \& {de Koter}(1997)}]{1997A&A...322..598S}
{Schaerer}, D., \& {de Koter}, A. 1997, \aap, 322, 598

\bibitem[{{Silich} {et~al.}(2005){Silich}, {Tenorio-Tagle}, \&
  {A{\~n}orve-Zeferino}}]{2005ApJ...635.1116S}
{Silich}, S., {Tenorio-Tagle}, G., \& {A{\~n}orve-Zeferino}, G.~A. 2005, \apj,
  635, 1116

\bibitem[{{Skillman}(1985)}]{1985ApJ...290..449S}
{Skillman}, E.~D. 1985, \apj, 290, 449

\bibitem[{{Torres-Peimbert} {et~al.}(1989){Torres-Peimbert}, {Peimbert}, \&
  {Fierro}}]{1989ApJ...345..186T}
{Torres-Peimbert}, S., {Peimbert}, M., \& {Fierro}, J. 1989, \apj, 345, 186

\bibitem[{{Townsley} {et~al.}(2006){Townsley}, {Broos}, {Feigelson}, {Brandl},
  {Chu}, {Garmire}, \& {Pavlov}}]{2006AJ....131.2140T}
{Townsley}, L.~K., {Broos}, P.~S., {Feigelson}, E.~D., {Brandl}, B.~R., {Chu},
  Y., {Garmire}, G.~P., \& {Pavlov}, G.~G. 2006, \aj, 131, 2140

\bibitem[{{T{\"u}llmann} {et~al.}(2008){T{\"u}llmann}, {Gaetz}, {Plucinsky},
  {Long}, {Hughes}, {Blair}, {Winkler}, {Pannuti}, {Breitschwerdt}, \&
  {Ghavamian}}]{2008ApJ...685..919T}
{T{\"u}llmann}, R., {Gaetz}, T.~J., {Plucinsky}, P.~P., {Long}, K.~S.,
  {Hughes}, J.~P., {et~al.} 2008, \apj, 685, 919

\bibitem[{{T{\"u}llmann} {et~al.}(2009){T{\"u}llmann}, {Long}, {Pannuti},
  {Winkler}, {Plucinsky}, {Gaetz}, {Williams}, {Kuntz}, {Pietsch}, {Blair},
  {Haberl}, \& {Smith}}]{2009ApJ...707.1361T}
{T{\"u}llmann}, R., {Long}, K.~S., {Pannuti}, T.~G., {Winkler}, P.~F.,
  {Plucinsky}, P.~P., {et~al.} 2009, \apj, 707, 1361

\bibitem[{{Wang} {et~al.}(1991){Wang}, {Hamilton}, {Helfand}, \&
  {Wu}}]{1991ApJ...374..475W}
{Wang}, Q., {Hamilton}, T., {Helfand}, D.~J., \& {Wu}, X. 1991, \apj, 374, 475

\bibitem[{{Wang} \& {Helfand}(1991)}]{1991ApJ...373..497W}
{Wang}, Q., \& {Helfand}, D.~J. 1991, \apj, 373, 497

\bibitem[{{Wang}(1999{\natexlab{a}})}]{1999ApJ...517L..27W}
{Wang}, Q.~D. 1999{\natexlab{a}}, \apjl, 517, L27

\bibitem[{{Wang}(1999{\natexlab{b}})}]{1999ApJ...510L.139W}
---. 1999{\natexlab{b}}, \apjl, 510, L139

\bibitem[{{Wang}(2004)}]{2004ApJ...612..159W}
---. 2004, \apj, 612, 159

\bibitem[{{Yang} {et~al.}(1994){Yang}, {Skillman}, \&
  {Sramek}}]{1994AJ....107..651Y}
{Yang}, H., {Skillman}, E.~D., \& {Sramek}, R.~A. 1994, \aj, 107, 651

\end{thebibliography}

\end{document}